\documentclass[mathleft]{an}
\usepackage{graphicx}
\usepackage{times}

\usepackage{amsmath}
\usepackage{amssymb}
\usepackage{natbib}
\bibpunct{(}{)}{;}{a}{}{,}

\def\be{\begin{equation}}
\def\ee{\end{equation}}
\def\bea{\begin{eqnarray}}
\def\eea{\end{eqnarray}}

\begin{document}

% The following seven commands are intended for editorial usage and should be ignored by
% the author(s).
\Pagespan{789}{}% Document's page range. 
% If second parameter is left empty, the last page is computed automatically.
\Yearpublication{2006}%
\Yearsubmission{2005}%
\Month{11}%   
\Volume{999}%  
\Issue{88}% 
% \DOI{This.is/not.aDOI}% 

\title{Roche volume filling and the dissolution of open star clusters}
\author{A. Ernst\inst{1}\fnmsep\thanks{email: aernst@ari.uni-heidelberg.de}, 
P. Berczik\inst{2,3,1}\fnmsep\thanks{email: berczik@mao.kiev.ua}, 
A. Just\inst{1}\fnmsep\thanks{email: just@ari.uni-heidelberg.de}, 
T. Noel\inst{1} 
}

\titlerunning{Dissolution of open star clusters}
\authorrunning{A. Ernst, P. Berczik, A. Just, T. Noel}
\institute{Astronomisches Rechen-Institut / Zentrum f\"ur Astronomie der Universit\"at Heidelberg, 
M\"onchhofstrasse 12-14,
69120 Heidelberg, Germany 
\and
National Astronomical Observatories of China, Chinese
Academy of Sciences, 20A Datun Rd., Chaoyang District, 100012, Beijing, China 
\and
Main Astronomical Observatory, National Academy of
Sciences of Ukraine, 27 Akademika Zabolotnoho St., 03680 Kyiv, Ukraine 
}

\received{...}
\accepted{...}

\label{firstpage}

\abstract{
From direct $N$-body simulations we find that the dynamical evolution of star clusters is strongly influenced by the Roche volume filling factor.
We present a parameter study of the dissolution of open star clusters with different Roche volume
filling factors and different particle numbers. 
We study both Roche volume underfilling and overfilling models and compare with
the Roche volume filling case. We find that in the Roche volume overfilling limit 
of our simulations two-body 
relaxation is no longer the dominant dissolution mechanism but the changing cluster potential. 
We call this mechnism ``mass-loss driven dissolution'' 
in contrast to ``two-body relaxation driven
dissolution'' which occurs in the Roche volume underfilling regime.  
We have measured scaling exponents of the dissolution time with the two-body relaxation time.
In this experimental study we find a decreasing scaling exponent 
with increasing Roche volume filling factor.  
The evolution of the escaper number in the Roche volume overfilling limit
can be described by a log-logistic
differential equation. 
We report the finding of a resonance condition which may play a role for the
evolution of star clusters and may be calibrated by the main periodic orbit in the large island of retrograde
quasiperiodic orbits in the Poincar\'e surfaces of section. We also report on the existence of a stability
curve which may be of relevance with respect to the structure of star clusters.
}

\maketitle

\raggedbottom

\section{Introduction}

%In contrast to globular clusters (GCs) which are 
%orbiting typically on eccentric orbits in the Galactic halo, open clusters (OCs) reside on near-%circular orbits in the 
%stellar disc of the Milky Way. 
While open clusters (OCs) dissolve, they are populating the disc with field
stars by continuous mass loss in the tidal field of the Milky Way.
The fraction of field stars which belonged once to a star cluster is estimated to be 10\% or
less \citep{Wielen1971, Miller1978},
up to 40\% \citep{Roeser2010}
or 100\% \citep{Maschberger2007}.
%%AJ

The feeding of the field population by OCs is determined by the initial cluster mass function (ICMF), the cluster formation rate (CFR) and the mass loss rates of clusters corrected for stellar evolution. The current ICMF can be directly determined by the young clusters of an unbiased cluster sample in the solar neighbourhood \citep{Kharchenko2013} or in nearby galaxies \citep{Boutloukos2003}. But for disentangeling the CFR and the cluster mass loss rates in the cluster distribution function $N(M,\tau)$ with mass $M$ and age $\tau$ additional information is needed. \citet{Lamers2005} used the dissolution timescales of \citet{Baumgardt2003} to analyse $N(M,\tau)$ in the LMC, M33, M51 and in the solar neighbourhood. In \citet{Lamers2005b} a more detailed analysis of the solar neighbourhood has shown that the present day CFR provides 30\% of the star formation rate as determined by \citet{Just2010}. It was also shown in \citet{Lamers2005b} that the dissolution timescales of OCs in the solar neighbourhood are significantly smaller than those predicted by \citet{Baumgardt2003}.

A major issue in the determination of cluster mass loss and lifetimes is the wide range of possible initial conditions. OCs form in giant molecular clouds and after a few Myr the volume is cleared from the interstellar medium. In this phase a large fraction of embedded clusters may become unbound and dissolve quickly (sometimes called ``infant mortality''). Since the dynamical state of the stellar component is strongly affected by the loss of the cloud potential and because the Galactic tidal field is not dominating the cluster formation process, the initial conditions of the isolated young clusters are not well constrained. The structure of the cluster may be characterized by three aspects, namely the dynamical state, the stellar content, and the density profile. After the gas removal, the OC may be supervirial and expand in a violent relaxation phase \citep[e.g.][]{Parmentier2012}. This phase lasts for a few crossing times and a compact core may survive, which then builds the starting point of the longterm evolution of the cluster. Therefore most investigations of the dynamical evolution of star clusters start in dynamical equilibrium and then add the tidal field. An initial mass segregation may alter the dynamical evolution of the cluster. Mass segregation is observed in some young, massive and compact clusters \citep[e.g.][]{Pang2013, Habibi2013}. Additionally, mass loss by stellar evolution depends strongly on the shape of the adopted initial mass function (IMF).
A pioneering ``survey'' of the dissolution of stellar clusters is the work of \citet{Fukushige1995}.
They investigated a parameter space of different concentrations and slopes of the IMF
and found that less concentrated clusters and/or those containing more massive stars are disrupted
sooner.

The density distribution of a cluster may be characterized by the core radius $r_c$, the half-mass radius $r_h$, and the cutoff radius $r_t$, where the density drops to zero. The general shape of the density profiles can be measured by the Lagrange radii $r_{i\%}$ containing $i\%$ of the cluster mass ($r_h=r_{50\%}$ and $r_t=r_{100\%}$). In the widely used King models \citep[lowered isothermal spheres, ][]{King1966} the concentration $c=\log(r_t/r_c)$ (or equivalently the depth of the potential well $W_0$) is a free parameter, which can be set to any positive number \citep[see][for more details]{Binney2008}. In contrast, the ratio of cutoff to half-mass  radius varies only by a factor of $\sim 3.3$ with the minimum at low concentration ($r_t/r_h=3$ for $W_0=1$) and a maximum at $W_0 \sim 8$ ($r_t/r_h=9.1$), which turns out to be a serious restriction for setting initial conditions of extended clusters.

The strength of the Galactic tidal field can be quantified by the Jacobi radius $r_J$, which is the distance of the Lagrange points $L_1$ and $L_2$, the saddle points of the effective potential, to the cluster centre.
The Jacobi radius for circular orbits is given by \citep{Kuepper2008, Just2009}
\be
r_J = \left[\frac{GM_{\rm cl}(r_J)}{(4-\beta_C^2)\Omega_C^2}\right]^{1/3}, \label{eq:rjcirc}
\ee
where $G$, $M_{\rm cl}(r_J)$, $\Omega_C$ and $\beta_C=\kappa_C/\Omega_C$ are the gravitational constant, 
the cluster mass inside $r_J$, the circular frequency and the ratio of epicyclic over circular 
frequency, respectively.
For clusters on a circular orbit (in an axi-symmetric Galactic potential) the Jacobi energy $E_J$ is a constant of motion and all stars inside the Roche volume given by $r_J$ with $E_J<E_{\rm J,crit}$, where $E_{\rm J,crit}$ is the effective potential at the Lagrange points $L_1$ and $L_2$, 
cannot escape. But there is no general criterion for a bound system.  \citet{Fellhauer2005} have shown that unbound, low density systems can survive for more than a Gyr in the tidal field. In general potential escapers with $E_J>E_{\rm J,crit}$ may stay a long time in the vicinity of the cluster, before they escape to the tidal tails or return to the cluster. \citet{Ross1997} derived a criterion for escape dependent on the offset of the guiding radius and the epicyclic radius of the star orbit. Applying this criterion to a flat rotation curve, the closest point of `safe' escape is at a distance of $\sim 2.6 r_J$. They have also shown that stars with arbitrarily large epicylic motion may return to the cluster. Therefore it is appropriate to count all stars inside $3 r_J$ to be bound 
as a simple criterion. 

For this study, we will use the 100\% Roche volume filling factor $\widehat{\lambda}=r_{100\%}/r_J$ to set the initial size of the cluster with respect to the tidal field. For practical reasons (see Section \ref{sec:method}) we will use  $\lambda'=r_{99\%}/r_J$ to scale the initial size of the cluster and determine $\widehat{\lambda}$ analytically. Since the  outer shells of the cluster contain only a small fraction of the cluster mass, the half-mass Roche volume filling factor $\lambda=r_h/r_J$ is a more robust measure of the impact of the tidal field on the cluster evolution.
Our goal is the analysis of the evolution of Roche volume overfilling star clusters.
There is a huge number of publications on numerical simulations concerning the dissolution of Roche volume filling or underfilling star clusters in tidal fields. We can refer only to a selected subset representing the main results relevant for our purpose.

%%AJ end
\citet{Engle1999} examined for the first time the evolution 
of Roche volume underfilling star cluster models
and found a relaxation driven expansion phase until the previously underfilling clusters
filled the Roche volume.
\citet{Engle1999} also presented for the first time a plot of lifetime vs.  
reciprocal Roche volume filling factor from direct $N$-body simulations 
and found increasing lifetime for decreasing filling factor.

\citet{Fukushige2000} calculated the Jacobi energy dependence of the time scale of
escape from a star cluster in a tidal field using a theoretical result of  \citet{MacKay1990}.

\citet{Baumgardt2001} used the calculation by \citet{Fukushige2000} and obtained the
scaling of the dissolution time with the two-body relaxation time for potential well
filling clusters in a tidal field. For the latter case he postulated a steady state 
equilibrium between escape 
and backscattering of potential escapers into the potential well and found that
the half-mass time scales with $\left[N/\ln(\gamma N)\right]^{3/4}$
in that equilibrium, where $\gamma$ is the factor in the Coulomb logarithm
occuring in the two-body relaxation time given by Eqn. (\ref{eq:t1}) below.
The potential escapers are stars which have been scattered above the 
critical Jacobi energy but which have not yet left the star cluster region.

\citet{Baumgardt2003} presented the results of a large parameter study of the
evolution of multi-mass star clusters in external tidal fields (i.e. a logarithmic halo). They used 
different particle numbers, orbital eccentricities and density profiles for star clusters, tending more
in the direction of globular clusters rather than towards the regime of OCs.

\citet{Tanikawa2005} simulated a comprehensive 
set of equal-mass star cluster models with different Roche volume filling factors
including for the first time Roche volume overfilling clusters. They cover a wide range of
particle numbers reaching the globular cluster regime and quantify how 
the dependence of the mass loss time scale on the two-body relaxation time  
scale depends sensitively on the strength of the tidal field as imposed by the Roche 
volume filling factor.

If two-body relaxation drives the evolution, 
the term proportional to $\left[N/\ln(\gamma N)\right]^{3/4}$ in the half-mass time according to the
theory in \citet{Baumgardt2001} can be well approximated by $BN^\eta$ with 
$\eta\approx 0.6$ \citep{Lamers2005}.
\citet{Lamers2005}, \citet{Lamers2005b}, \citet{Gieles2008} and \citet{Lamers2010}
find that the dissolution time (e.g. half-mass time) scales with
$\left(M_{\rm cl}/M_\odot\right)^{0.6-0.8}$ for the Roche volume filling case, where
$M_{\rm cl}$ is the initial cluster mass
and the exponent
depends on the parameter $W_0$ of the \citet{King1966} initial model. 

This particularly means that the equilibrium postulated by \cite{Baumgardt2001} may not be realized. 
Initially this can be in fact true, as Baumgardt himself notes. The reason is that the potential 
escaper regime may be initially overpopulated. The dependence on $W_0$ noted above is a hint that this
scenario plays a role.
\citet{Baumgardt2001} did not state the linear stability
analysis of the equilibrium which he postulated in his 2001 paper.
The proof that the postulated steady state equilibrium is attracting
typical non-equilibrium initial states and the quantification of the
attraction strength and time scale are still open issues.

In the present study, we will present numerical evidence for the fact that open star clusters 
in the Roche volume overfilling regime dissolve mainly due to the changing cluster potential and the shear forces 
of the differentially rotating galactic disk. 
We call this mechanism ``mass-loss driven dissolution'' 
in contrast to the ``two-body relaxation driven dissolution'' which occurs from the 
Roche volume underfilling regime up to the 
Roche volume filling case \citep[see also][based on simpler models]{Whitehead2013}.

We concentrate in the present study on the properties and evolution of classical OCs which already left 
their parent molecular cloud. 
Therefore the formation process and the early phase of gas expulsion
of OCs is beyond the scope of the project. 
On the other hand, mass loss of the OCs due to stellar evolution (supernovae, stellar winds, planetary
nebulae) will be taken into account.
We systematically study the influence of the Roche volume 
filling factor which we have chosen as the main free parameter.
In this sense, the present study aims to extend the study by \citet{Engle1999, Baumgardt2003, Tanikawa2005}
into the overfilling regime.
In particular, we aim at obtainig scaling 
exponents of the dissolution time with the two-body relaxation time scale
and at deriving a fitting formula for the dissolution time (half-number time).

This paper is organized as follows: In Section 2 we shortly explain the method of direct $N$-body
simulations in an analytic background potential and the programs {\sc nbody6tid} and 
$\varphi$-{\sc grape+gpu}. 
In Section 3 we discuss the parameter space.
Section 4 contains the results and Section 5 the conclusions.

\section{Method}

\label{sec:method}

\begin{table}
\caption{The list of galaxy component parameters.}
\label{tab:gal-par}
\begin{center}
\begin{tabular}{lcrr}
\hline\noalign{\smallskip}
Component & M [M$_\odot$] & $a~[{\rm kpc}]$ & $b~[{\rm kpc}]$ \\
\noalign{\smallskip}
\hline
\noalign{\smallskip}
 Bulge & $1.4 \times 10^{10}$ & 0.0 &  0.3 \\
 Disk  & $9.0 \times 10^{10}$ & 3.3 &  0.3 \\
 Halo  & $7.0 \times 10^{11}$ & 0.0 & 25.0 \\ 
\hline
\end{tabular}
\end{center}
\end{table}

The dynamical evolution of OCs is calculated as an $N$-body problem 
in an analytic background potential of the Milky Way. 
For the background Milky Way potential, we use the same model as in \citet{Kharchenko2009,
Just2009, Ernst2010, Ernst2011},
i.e. an axisymmetric three-component model, 
where the bulge, disk, and halo are described by Plummer-Kuzmin models 
\citep{Miyamoto1975} with the potential

\be
\Phi(R,z) = - \frac{ GM }{ \sqrt{R^2 + (a + \sqrt{b^2 + z^2} )^2} }. \label{eq:eq-gal}
\ee

\noindent
The parameters $a, b$, and $M$ of the Milky Way model are given in Table \ref{tab:gal-par}
for the three components. For details of the rotation curve, tidal field and the saddle points
of the effective potential see \citet{Just2009,Ernst2010}.

%%AJ

We cover a large range in particle number $N$ and Roche volume filling factor $\widehat{\lambda}$. For the analysis of the mass loss we take the average over sets of random realisations in order to reduce the impact of random noise. For details see Section \ref{sec:par}.
As the main parameter to measure the dissolution time we use the half-number-time $t_{\rm 50}$, where 50\% of the initial particles are lost.

%%AJ end
For the solution of the $N$-body problem the $N$-body programs {\sc nbody6tid} and
$\varphi$-{\sc grape+gpu} were used. 

 For this study, we use $\lambda'$, the 99\% Roche
volume filling factor, as a measure for the Roche volume filling
since the $99$\% Lagrange radius is a statistically more 
robust measure than the $100$\% Lagrange radius. 
 For \citet{King1966} models, which have a cutoff radius where the density drops to zero, 
the ratios between $99$\% and $100$\% Lagrange radius are fixed.
The conversion factor between $r_{99\%}$ and  $r_{100\%}$
is 1.521 for a $W_0=6$ King model. The conversion factor between $r_{h}$ and  $r_{99\%}$
is 4.486 for a $W_0=6$ King model. 

After the random realization of positions, velocities and stellar masses 
the Jacobi radius $r_J$ and the Lagrange radii $r_{i\%}$ were determined, the cluster size was scaled to realize the selected Roche volume filling factors $\lambda'$ in the tidal field (Table \ref{tab:parspace}), before the simulation was started.

\subsection{{\sc nbody6tid} code}

Originally, the program {\sc nbody6tid} was written for Galactic centre studies 
\citep{Ernst2009a, Ernst2009b} and called {\sc nbody6gc}. Later the three-component 
Plummer-Kuzmin Milky Way model 
based on Eqn. (\ref{eq:eq-gal}) and Table \ref{tab:gal-par} was added to treat the tidal field
similar to $\varphi$-{\sc grape+gpu} (see below), and the program
was renamed {\sc nbody6tid}.
In {\sc nbody6tid}, the galactic centre position is modelled as a pseudo-particle carrying the Galactic potential in an orbit 
around the star cluster, which must be located close to the origin of coordinates.
In addition to the equations of motion of the star cluster $N$-body problem in the comoving coordinate system, which are solved
using a fourth-order Hermite integration scheme \citep{Makino1992} with individual 
(hierarchical) block  time steps \citep{Aarseth2003}, {\sc nbody6tid} solves the equations of motion for the galactic centre orbit 
with a time-transformed eighth-order composition scheme 
\citep{Yoshida1990, McLachlan1995, Mikkola1999, Preto1999, Mikkola2002}. 
The tidal force of the background potential acts on all particles in the 
$N$-body system. It is added to the regular 
force part of {\sc nbody6} \citep{Ahmad1973, Aarseth2003}.
Furthermore, the tidal force is added as a perturbation to the KS 
regularization \citep{Kustaanheimo1965} of {\sc nbody6} 
\citep{Aarseth2003}. The time derivative of tidal acceleration (``jerk'') is 
also calculated
and added appropriately in the fourth-order Hermite integration scheme.
Also, the Chandrasekhar dynamical friction force \citep{Chandrasekhar1943, Binney1987, Binney2008} 
is implemented using an implicit midpoint method \citep{Mikkola2002}.
Stellar evolution is modelled with the fitting formulas of \citet{Hurley2000}.
Kicks for compact objects are applied in the {\sc nbody6tid} runs. 
For stellar mass black holes and neutron
stars the kick velocities are drawn from a Maxwellian with a 1D dispersion
which equals twice the velocity unit \citep{Heggie1986} in km s$^{-1}$
% \citep{Hansen1997} 
while for the white dwarfs the corresponding 
1D dispersion is chosen to be $5$ km s$^{-1}$ \citep[e.g.][for a lower bound]{Fellhauer2003}.
For high particle numbers $N= 20k, 50k$ the serial {\sc GPU}
variant {\sc nbody6tidgpu} is partly used in the present study, based on {\sc nbody6gpu} 
\citep{Nitadori2012}, a version of {\sc nbody6} \citep{Aarseth2003}
which uses NVIDIA type GPU's with CUDA\footnote{\tt http://www.nvidia.com} 
library support for the force and jerk evaluations.

\subsection{$\varphi$-{\sc grape+gpu} code}

We also show a few models calculated with the
$\varphi$-{\sc grape+gpu} $N$-body code which also uses the 
fourth-order Hermite integration scheme \citep{Makino1992}
with individual (hierarchical) block time steps and includes the three-component
Plummer-Kuzmin model based on Eqn. (\ref{eq:eq-gal}) and Table \ref{tab:gal-par}. 
The external gravity part for the tidal field
is calculated in the galactocentric reference frame. Against rounding errors the $N$-body
problem of the star cluster is solved in the local cluster frame as in {\sc nbody6tid}.
The first version of the code was written from scratch 
in {\sc C} and originally designed to use the 
{\sc grape6a} clusters for the $N$-body task integration 
\citep{Harfst2007}. In the present version of the $\varphi$-{\sc grape+gpu} code we 
use the NVIDIA type GPU's with CUDA 
library support, with the external {\sc sapporo}  
library \citep{Gaburov2009}, which emulates for us the standard 
{\sc grape6a} library calls on the NVIDIA GPU hardware.
The $\varphi$-{\sc grape+gpu} code was extensively tested 
and already long time successfully used in our earlier
Milky Way star cluster dynamical mass loss simulations: 
\citet{Just2009,Kharchenko2009,Ernst2010}.\footnote{The first 
original public version of the $\varphi$-{\sc grape+gpu} 
code can found here: \\
{\tt ftp://ftp.mao.kiev.ua/pub/users/berczik/phi-GRAPE+GPU/}
}
The stellar evolution treatment of $\varphi$-{\sc grape+gpu} is in detail described
in \citet{Kharchenko2009}. For the $\varphi$-{\sc grape+gpu} runs in the present 
study kicks for compact objects have not been applied for technical reasons.

\section{Parameter space} \label{sec:par}

%Nomenclature:
% A = O_1  lambda' = 1
% D = O_2   lambda' = 2
% E = O_3  lambda' = 3
% B = F     lambda' = 2/3
% C = U_2    lambda' = 1/3
% G = U_1    lambda' = 1/2

\begin{table}
\caption{Overview of the parameter space and of the ensembles. Each series comprises
a set of ensembles with different particle numbers $N$, {\sc nbody6tid}
ensemble sizes $n$ or $\varphi$-{\sc grape+gpu} ensemble sizes $n'$ . 
The quantities $r_{99\%}/r_J$,
$W_0$ and $R_g$ are the Roche volume filling factor, the King parameter and the
galactocentric radius, respectively.
The letter ``k'' stands for ``kilo'' (=1000). A ``+'' or a ``*'' means that the 
corresponding ensemble has been calculated with {\sc nbody6tid}
or with $\varphi$-{\sc grape+gpu}, respectively. 
A ``-'' means that the corresponding ensemble
has not been calculated. The number in round brackets
after a ``*'' denotes the number of $\varphi$-{\sc grape+gpu} simulations, i.e. the 
$\varphi$-{\sc grape+gpu} ensemble size $n'$.
}
%\label{tab:parspace}
\begin{center}

\begin{tabular}{lcccccccc}
\hline
Series & $F$ & $U_1$ & $U_2$ & $O_1$ & $O_2$ & $O_3$ \\
\hline
$W_0$ & 6 & 6 & 6 & 6 & 6 & 6\\
$\lambda=r_{h}/r_J$  & 0.149 & 0.111 & 0.074 & 0.222 & 0.446 & 0.669   \\
$\lambda'=r_{99\%}/r_J$ & 2/3 & 1/2 & 1/3 & 1 & 2 & 3  \\
$\widehat{\lambda}=r_{100\%}/r_J$ & 1 & 3/4 & 1/2 & 3/2 & 3 & 9/2 \\
$R_g$ [kpc] & 8 & 8 & 8 & 8 & 8 & 8 \\
\hline
$(N, n)$ & $F$ & $U_1$ & $U_2$ & $O_1$ & $O_2$ & $O_3$ \\
\hline
$(50, 512)$ & + & + & + & + & + & + \\
$(100, 256)$ & + & + & + & + & + & + \\
$(200,128)$ & + & + & + & + & + & + \\
$(500, 64)$ & + & + & + & + & + & + \\
$(1\mathrm{k}, 32)$ & + & + & + & + & + & + \\
$(2\mathrm{k}, 16)$ & + & + & + & + & + & + \\
$(5\mathrm{k}, 8)$ & + & + & + & + & + & + \\ 
$(10\mathrm{k}, 4)$ & + & + & + & + & + & + \\
$(20\mathrm{k}, 2)$ & + & - & * (3) & +* (3) & +* (3) & +* (6) \\
$(50\mathrm{k}, 1)$ & - & - & -  & +* (1) & * (3) & * (6)  \\
$(100\mathrm{k}, -)$ & - & - & -  & - & * (6)  & * (6) \\
$(1000\mathrm{k}, -)$ & - & - & -  & * (1) & - & - \\
\hline
\end{tabular}
\end{center}
\label{tab:parspace}
\end{table}

Table \ref{tab:parspace} shows an overview over the parameter space.
We employ ensemble averaging over $n$ 
{\sc nbody6tid} simulations or $n'$ $\varphi$-{\sc grape+gpu} simulations.

We have chosen the minimum particle number to be $N=50$. We remark that the 
definition of the two-body relaxation time in Eqn. (\ref{eq:t1}) below 
breaks down for small $N$ due to the
minimum in $N/\ln(\gamma N)$ at $N=136$ for $\gamma=0.02$. The 
evolution of the system is no longer governed by small-angle scatterings in such a small-$N$ regime.

All models were  \citet{King1966} models with the King parameter $W_0=6$ placed on
a circular orbit at the Galactocentric radius $R_g=8$ kpc in $z=0$ plane of the 
Galactic tidal field based on Eqn. (\ref{eq:eq-gal}) and Table \ref{tab:gal-par}. In all models, we 
applied a \citet{Kroupa2001} initial mass function with $0.08 < m/M_\odot < 100.00$,
where $m$ is the stellar mass. We remark that the IMF can have a drastic effect on the 
lifetimes of star clusters as \citet{Engle1999} demonstrated numerically.
For the metallicity we chose the solar metallicity $Z=0.02$ in
all models. We remark that this value may not be up-to-date anymore. However, it
allows for comparisons with older simulations. 

The lower part of Table \ref{tab:parspace}  shows an overview over the simulations which have
been carried out. A ``+'' or a ``*'' means that the corresponding ensemble has been 
calculated with {\sc nbody6tid} or with $\varphi$-{\sc grape+gpu}, respectively. 
%A ``-'' means that the corresponding ensemble has not been calculated.
The ensembles marked with  a ``-'' sign have  not been simulated.
The number in round brackets
after a ``*'' denotes the number of additional $\varphi$-{\sc grape+gpu} simulations, i.e. the 
$\varphi$-{\sc grape+gpu} ensemble size $n'$.

We remark that \citet{Baumgardt2003} investigated the Roche volume filling case
with $\widehat{\lambda} = 1$ corresponding to our series F.

\subsection{{\sc nbody6tid} parameter space} 

The following parameter space was adopted for the {\sc nbody6tid} simulations:
For the ordered pairs $(N,n)$ we choose $(50,512)$, $(100,256)$, $(200,128)$, $(500,64)$, $(1000,32)$,
$(2000,16)$, $(5000,8)$, $(10000,4)$, $(20000,2)$, $(50000,1)$, where $N$ is the 
particle number and $n$ the number of {\sc nbody6tid} runs per ensemble. 
For each ordered pair $(N,n)$ we computed 6 ensembles
corresponding to the Roche volume filling factors $\lambda'=r_{99\%}/r_J= 1/3 \ (U_2), 1/2 \ (U_1), 2/3 \ (F), 1 \ (O_1), 2 \ (O_2) , 3 \ (O_3)$.
The full {\sc nbody6tid} parameter space comprised approx. $50$ ensembles 
or approx. $6000$ runs, respectively.

\subsection{$\varphi$-{\sc grape+gpu} parameter space}

The $\varphi$-{\sc grape+gpu} simulations were computed additionally.
The main reason to use both (similar) $N$-body codes was, that we
wanted to compare the results of both programs and
to apply the advantages of both programs to the same star cluster
dynamical evolution scientific application. 
The $\varphi$-{\sc grape+gpu} simulations include one model with $N=1$ million 
for series $O_1$.

\section{Results}

%%AJ

\subsection{Random scatter}

\begin{figure}
\includegraphics[angle=90,width=0.5\textwidth]{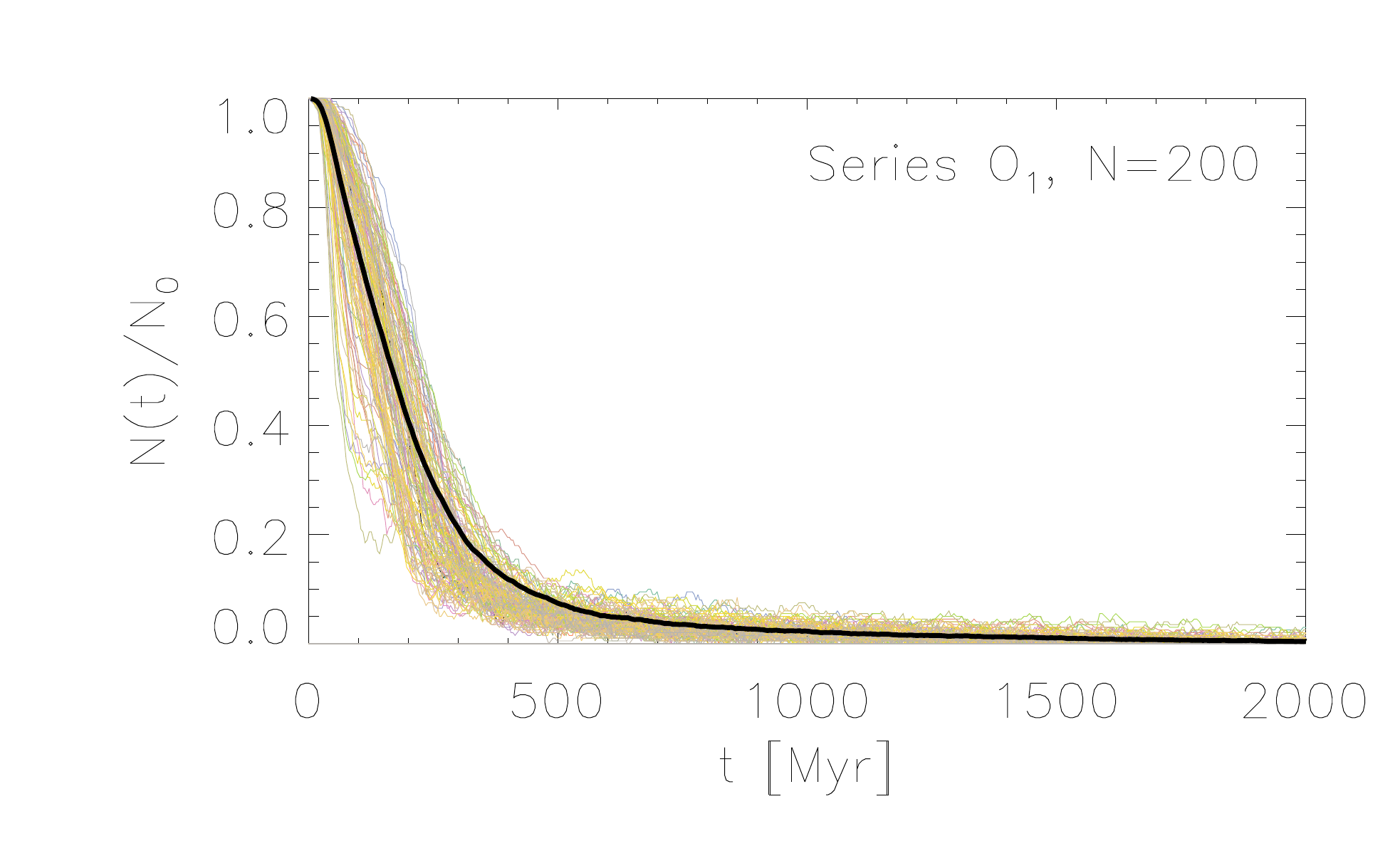} 
\caption{Example for the evolution of the particle number, ensemble with $(N,n)=(200,128)$.
The ensemble mean is marked by the thick solid black line. 
} 
\label{fig:nt200_A}
\end{figure}

\begin{figure}
\includegraphics[angle=0,width=0.5\textwidth]{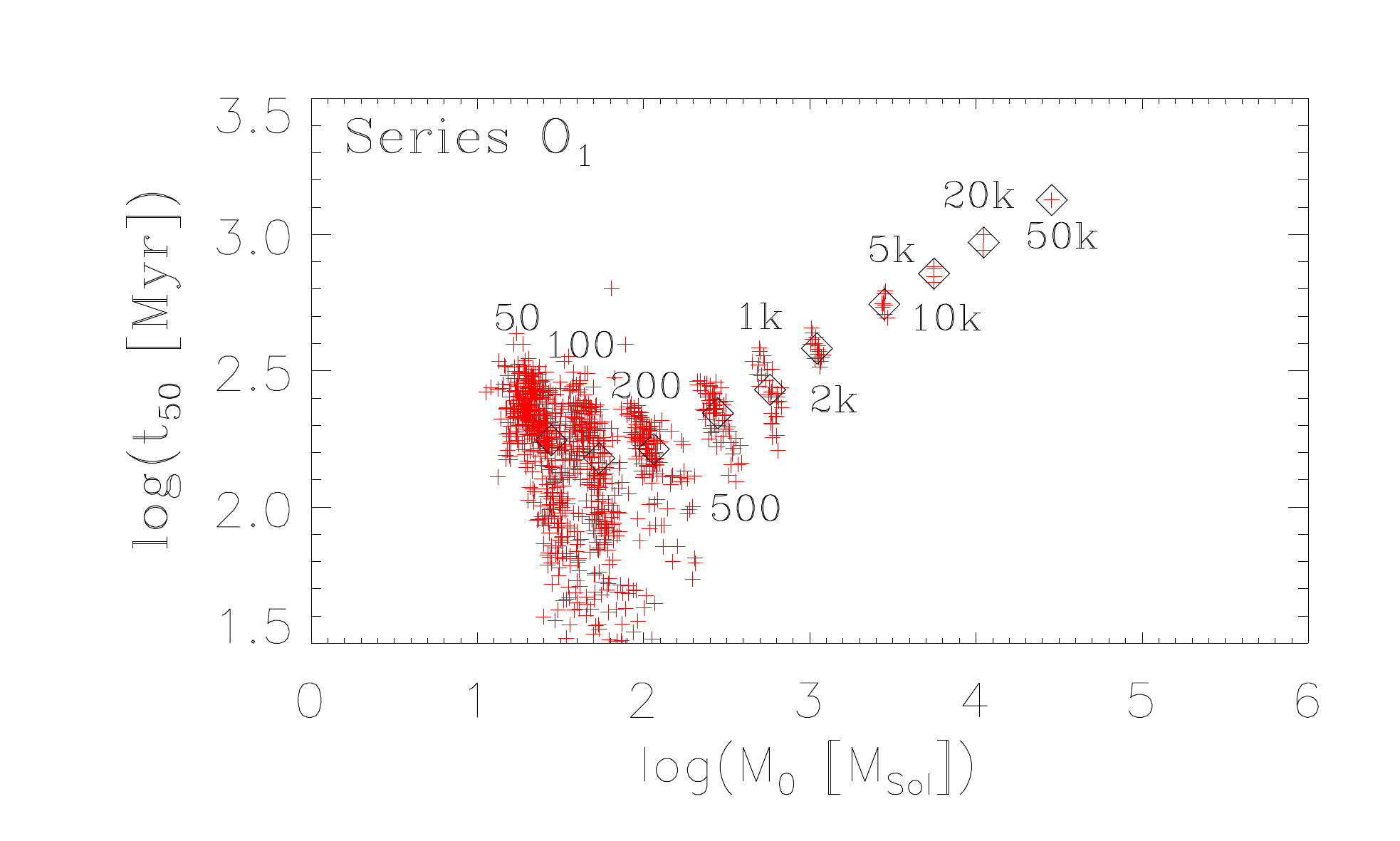} 
\caption{Half-number times versus initial mass for all runs of Series $O_1$. The particle
numbers are marked by labels. The mean values are marked by diamonds.
} 
\label{fig:t50all_A}
\end{figure}

\begin{figure}
\includegraphics[angle=0,width=0.5\textwidth]{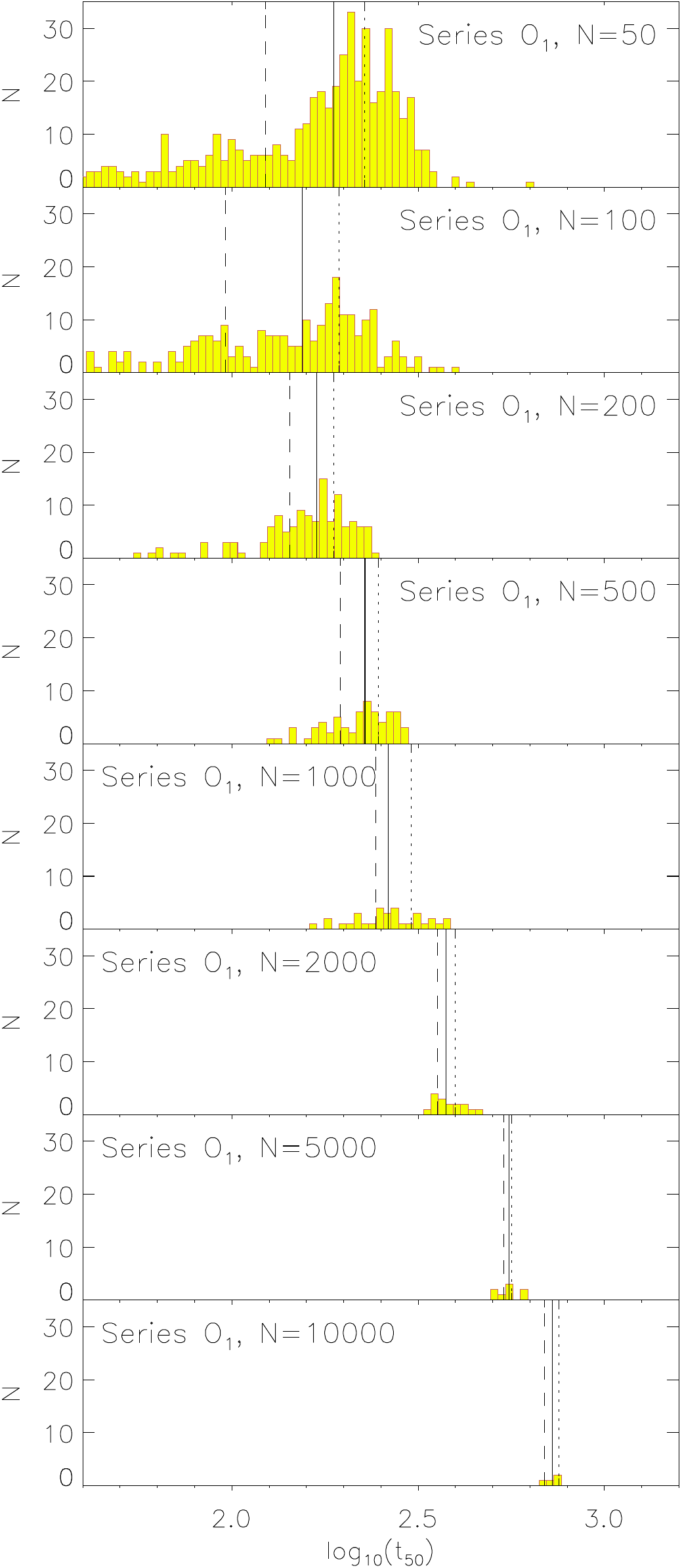} 
\caption{Distributions of half-number times $t_{50}$ for the low-$N$ ensembles of series $O_1$.
The binsize is $0.02$ dex in $\log_{10}(t_{50})$.
The ensemble median $Q_{50}$ is marked by the solid black line. The dotted and dashed lines mark the 
$30$\% and $70$\% quantiles $Q_{30}$ and $Q_{70}$ of the corresponding distribution, respectively.} 
\label{fig:t50hists3}
\end{figure}

For a cluster with $N=1000$ or smaller there is a considerable scatter in the realized cluster mass and size due to the random population of the high mass end of the IMF and of the outer shells of the cluster. 
Figure \ref{fig:nt200_A} shows the time evolution of all $128$ runs of the ensemble with $N=200$
of Series $O_1$. The ensemble mean is marked by the thick solid black line.
There is a strong scatter in mass loss times.
Figure \ref{fig:t50all_A} shows the half-number times $t_{50}$ as function of initial mass for all $(N,n)$ ensembles of series $O_1$. For each fixed $N$ there is a clear anticorrelation with initial mass showing that a few high-mass stars accelerate the dissolution significantly. Figure \ref{fig:t50hists3} shows the corresponding distributions of $t_{50}$ for the low-$N$ ensembles of series $O_1$. The binsize is $0.02$ dex in $\log_{10}(t_{50})$.
The ensemble median $Q_{50}$ is marked by the solid black line. The dotted and dashed lines mark the 
$30$\% and $70$\% quantiles $Q_{30}$ and $Q_{70}$ of the corresponding distribution.
The {\sc coyote} library in {\sc idl} \citep{Fanning2011} has been used.

%%AJ end

\begin{figure}
\includegraphics[angle=90,width=0.5\textwidth]{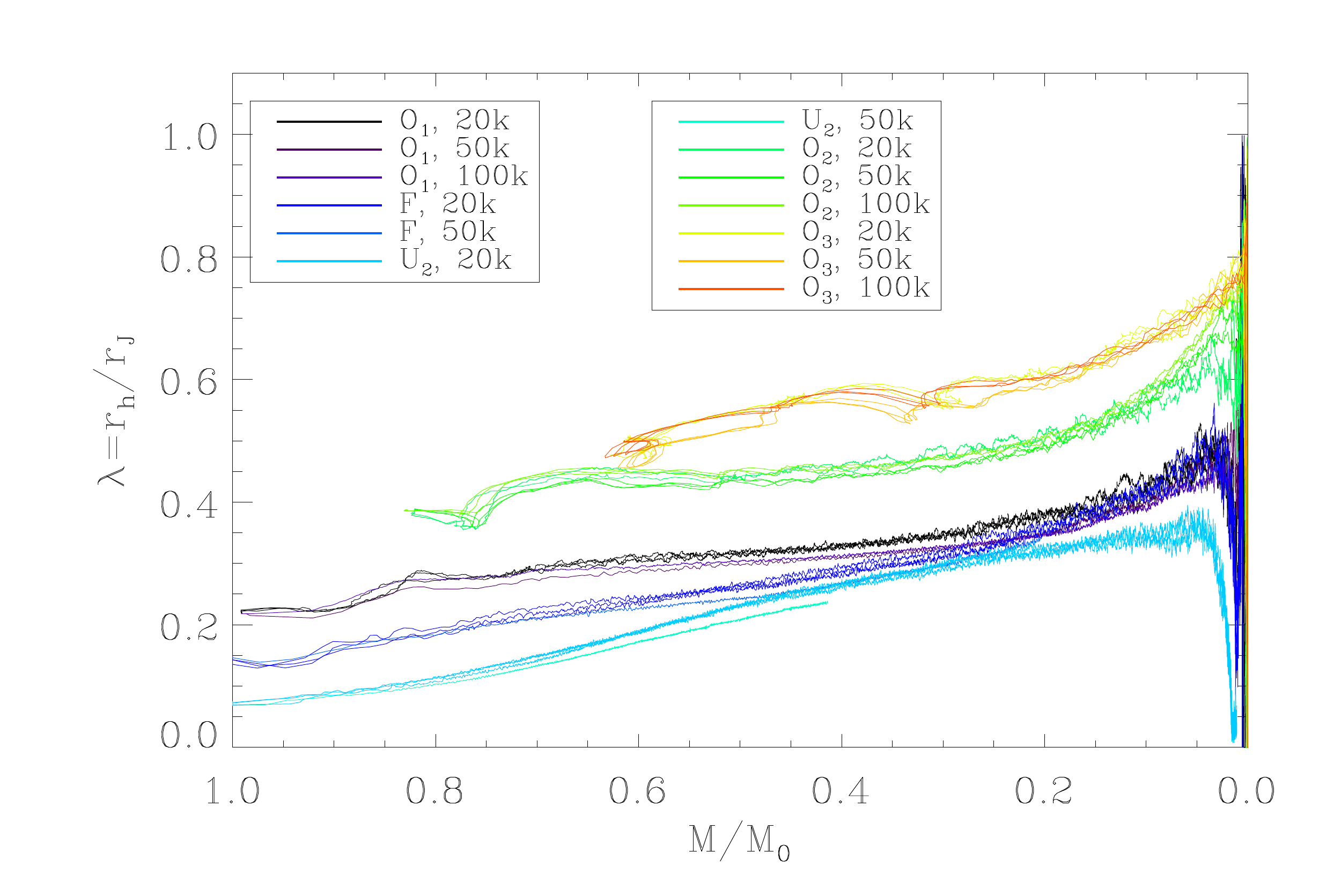} 
\caption{Evolution of the half-mass Roche volume filling factor $\lambda$ 
as a function of the bound mass $M/M_0$.
} 
\label{fig:lambdatall}
\end{figure}

%%AJ
In Figure \ref{fig:lambdatall} we compare the evolution of the half-mass radius as function of
 relative mass loss $M(t)/M_0$ in terms of $\lambda$ for different series. This shape parameter $\lambda$ increases during the evolution similar to the result of \citet{Fukushige1995} and it is independent of $N$. There is a clear separation of the different series showing that there is a memory of the initial relative size of the half-mass radius. 
%In the underfilling clusters a compact core of stellar black holes forms. Therefore $\lambda$ drops strongly at %the end of the evolution.
%%AJ end

\subsection{Mass loss}

\begin{figure*}
\includegraphics[angle=90,width=0.45\textwidth,height=0.9\textheight]{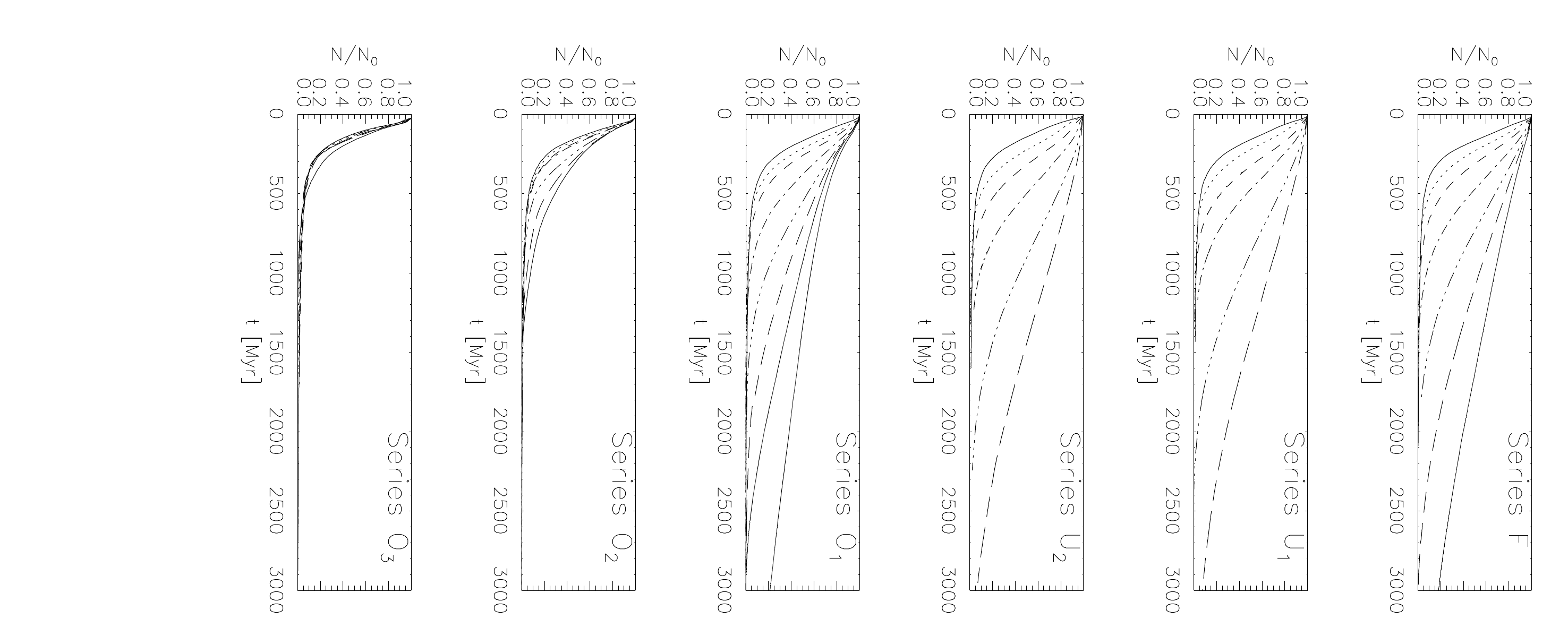} 
\includegraphics[angle=90,width=0.5\textwidth,height=0.9\textheight]{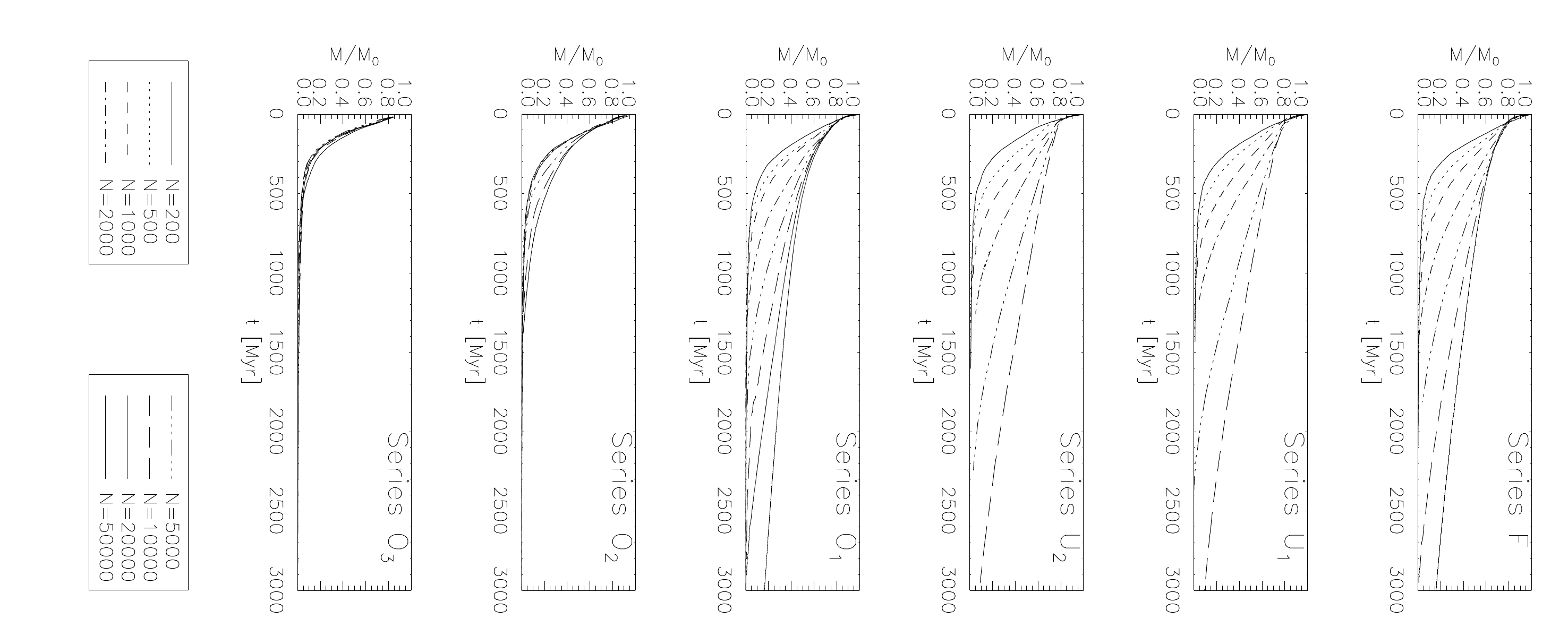} 
\caption{Evolution of the particle number within three times the Jacobi radius 
(left panels) and cluster mass within three times the Jacobi radius (right panels). 
From left to right in each panel: $N=200, 500, 1k, 2k, 5k, 10k, 20k, 50k$.
The sharp initial decrease in the total mass is due to the stellar evolution mass loss.}
\label{fig:meanntmtall}
\end{figure*}

The left-hand side of Figure \ref{fig:meanntmtall} shows the time evolution of the arithmetic ensemble mean of the normalized 
particle number $N(t)/N_0$ within three times the Jacobi radius, which we defined to be bound.

All curves decline monotonically as the simulated star clusters lose mass.
The moderately Roche volume underfilling series $U_1$ and the 
strongly Roche volume underfilling series 
$U_2$ dissolve slower than the Roche volume filling series $F$. The reason is that
they first need a phase of expansion to fill the Roche volume \citep[e.g.][]{Engle1999}.
It can be seen that in the moderately Roche volume overfilling
series $O_2$ the $N$-dependence is much weaker than in the Roche volume overfilling series $O_1$. 
Moreover, in the strongly Roche volume overfilling series $O_3$ the $N$-dependence has almost completely vanished.
This indicates that two-body relaxation is not responsible for the dissolution.

The right-hand side of Figure \ref{fig:meanntmtall} shows the time evolution of the 
arithmetic ensemble mean of the normalized 
mass $M(t)/M_0$ within three times the Jacobi radius.
All curves show a strong initial decrease due to the stellar evolution mass loss.
The half-mass times are significantly shorter than the half-number times due to the
stellar evolution mass loss. 
%%AJ
As a consequence the binding energy and Roche volume decrease, which depends on the cluster mass, faster than 2-body relaxation, which depends on the number of stars. 
%%AJ end
These differences decrease with increasing Roche volume
filling factor.

\subsection{Dissolution times}

There are four relevant time scales involved in the dissolution of star clusters in the Galactic tidal 
field: (i) The two-body relaxation time $t_{\rm rx}$, (ii) crossing time $t_{\rm cr}$, (iii) orbital  time
$t_{\rm orb}$ and (iv) stellar evolution time $t_{\rm stev}$. The first three
of them scale as

\bea
&&t_{\rm rx} \propto \frac{N}{\ln (\gamma N)} t_{\rm cr}, \label{eq:t1} 
\eea
\bea
&&t_{\rm cr} \propto \left(\frac{GM_{\rm cl}}{r_J^3}\right)^{-1/2} \lambda^{3/2}\propto 
t_{\rm orb}\lambda^{3/2}, \label{eq:t2}
\eea
\bea
&&t_{\rm orb} \propto \left(\frac{GM_{\rm g}}{R_g^3}\right)^{-1/2},\label{eq:t3}
\eea

\noindent
where $M_g$, $R_g$, and $\gamma$,
are the the enclosed Galaxy mass, the galactocentric radius, 
and the factor in the Coulomb logarithm 
\citep{Giersz1994, Giersz1996}, respectively. $r_J$ is given by
Eqn. (\ref{eq:rjcirc}). We note that we use the initial 
two-body relaxation time $t_{\rm rx}$ throughout this study and not the current one. 
%%AJ
The stellar evolution time $t_{\rm stev}$ depends only on the IMF and the metalllicity, which are fixed in our study.
%%AJ end
For ensembles with large particle numbers $N$ and correspondingly large 
two-body relaxation times $t_{\rm rx}$ the mass loss due to stellar evolution is clearly
distinguishable from the two-body relaxation driven evolution and becomes important with respect to it
as can be seen in right-hand side
panels of Figure \ref{fig:meanntmtall}.

\begin{figure*}
\centering
\includegraphics[angle=90,width=0.9\textwidth]{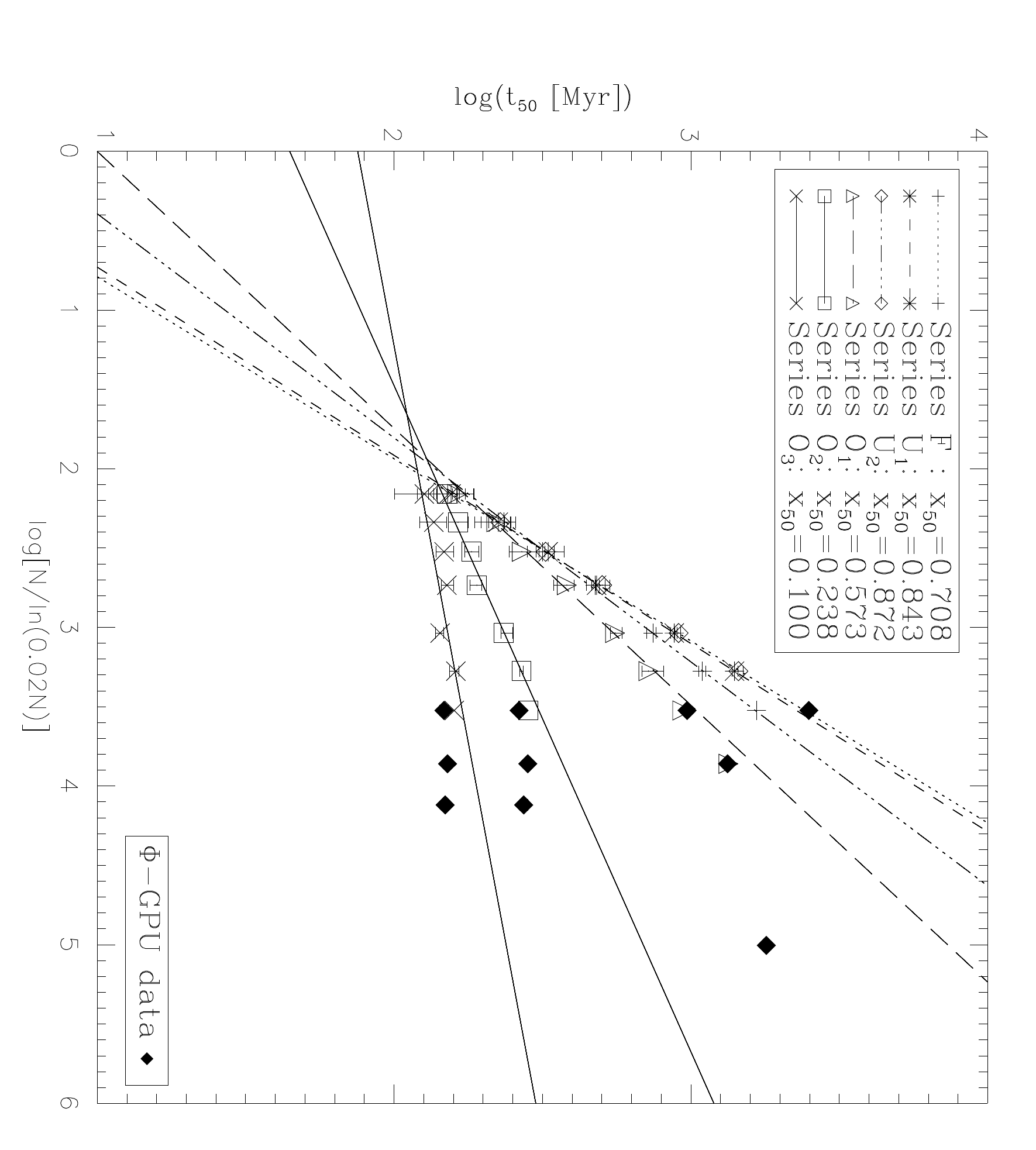} 
\caption{%Top panel: 
Half-number times in Myr as functions of $N/\ln(0.02 N)$ (dots) and straight-line
fits, each using the six lowest-$N$ data points (for $N=200, 500, 1\mathrm{k}, 2\mathrm{k}, 5\mathrm{k}, 10\mathrm{k}$) to obtain the slopes in the low-$N$ OC limit. The results of a 
few $\varphi$-{\sc grape+gpu} simulations are also shown. The $\varphi$-{\sc grape+gpu}
simulation with $1$ million particles belongs to series $O_1$. The weights and error bars 
are calculated from the quantiles $Q_{30,N}$, $Q_{50,N}$ and $Q_{70,N}$ (see text and Eqns. (\ref{eq:fiterror1}) - (\ref{eq:fitweights})).
%Bottom panel: Half-number times in Myr as functions of $\lambda'=r_{99\%}/r_J$.
} 
\label{fig:thnall}
\end{figure*}

\hspace{-1cm}
\begin{table}
\caption{The slopes $x_{50}$.}
\label{tab:thnall}
\begin{center}
\begin{tabular}{llcccccc}
\hline\noalign{\smallskip}
Ser. & $\gamma$ & $F$ &  $U_1$ &  $U_2$ & $O_1$  & $O_2$ & $O_3$ \\
$x_{50}$ & 0.02 & 0.708 & 0.843 & 0.872 & 0.573 & 0.238 & 0.100 \\
$x_{50}$ & 0.11 & 0.630 & 0.752 & 0.764 & 0.494 & 0.212 & 0.100 \\
\hline
\end{tabular}
\end{center}
\end{table}

Figure \ref{fig:thnall} shows that, in the low-$N$ regime of OCs,
the half-number time $t_{50}$ scales directly
with a power $x_{50}(\gamma,\lambda')$ of the two-body relaxation time $t_{\rm rx}$.
Figure \ref{fig:thnall} also shows the corresponding straight-line 
fits for the determination of $x_{50}(\gamma,\lambda')$. 
The 6 lowest-$N$ data points of the {\sc nbody6tid} ensembles 
have been used for the fitting of power laws.
 For the least-squares-fitting, we used the {\sc mpfit} package in
{\sc idl} \citep[for the Levenberg-Marquardt algorithm]{Markwardt2009, More1978}. 

\begin{figure}
\includegraphics[angle=0,width=0.5\textwidth]{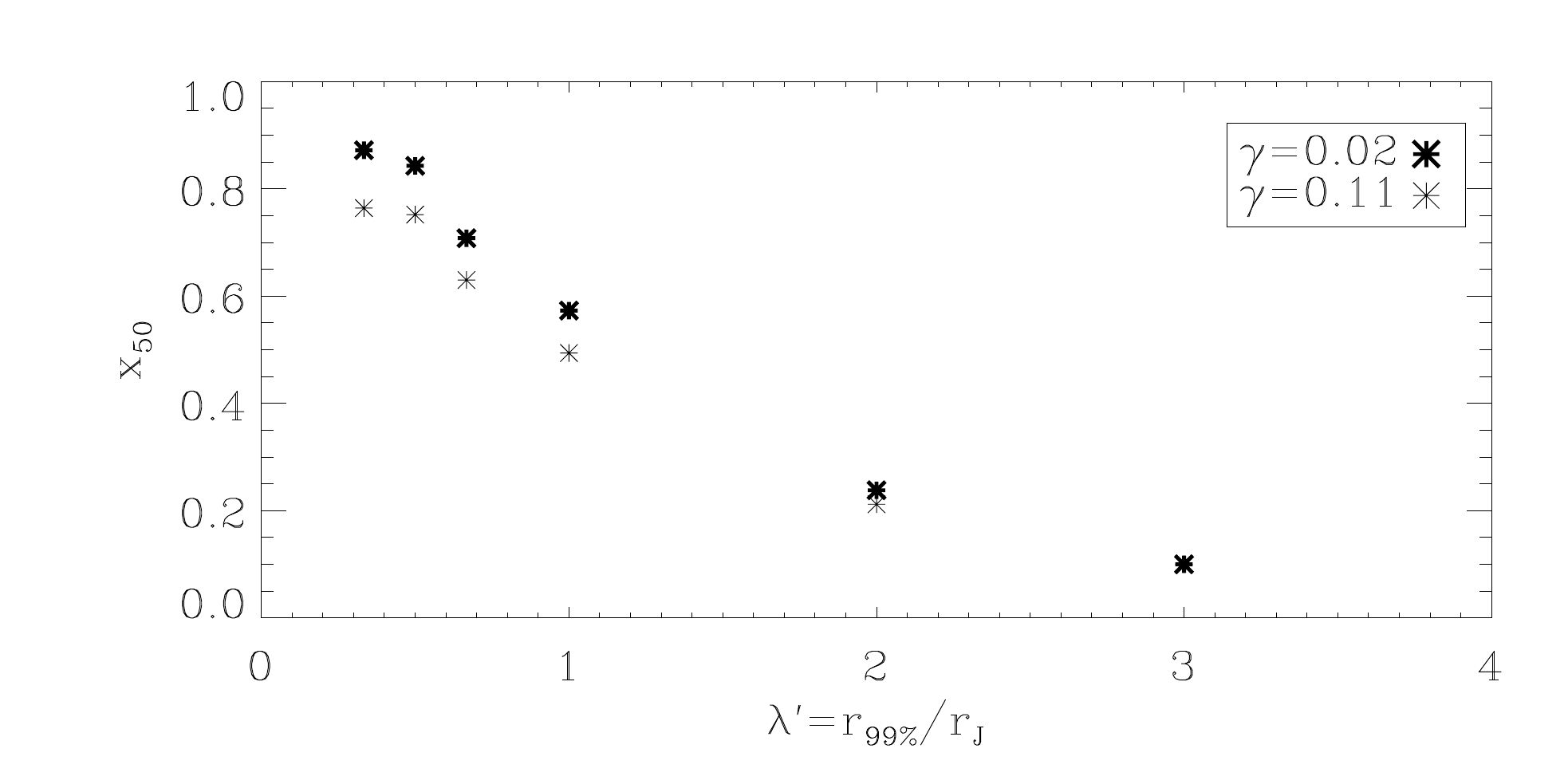} 
\caption{Exponent $x_{50}$ from Eqn. (\ref{eq:ansatz}) as a function of $\lambda'=r_{99\%}/r_J$.
Thick dots: $\gamma=0.02$; thin dots: $\gamma=0.11$.} 
\label{fig:exp}
\end{figure}

We find in this study for the half-number time $t_{50}$ the approximate expression

\be
\frac{t_{50}}{T}  \approx \left[\frac{1}{C}\frac{N}{\ln(\gamma N)}\right]^{x_{50}(\gamma, \lambda')} 
\label{eq:ansatz}
\ee

\noindent
with $T \approx 125 \ \mathrm{Myr} \propto t_{\rm orb}$ and $C = 80-100$. The exponents 

\be
x_{50}=\frac{d\log t_{50}}{d\log t_{\rm rx}}
\ee 

\noindent
for the half-number times are given in the legend of Figure \ref{fig:thnall} and Table \ref{tab:thnall},
and they are plotted in Figure \ref{fig:exp} against the Roche volume filling factor.

The upper and lower errors in Figures \ref{fig:thnall}, \ref{fig:tifitex205080}  
and \ref{fig:xiallq2} are given by 

\bea
\Delta^+ \log_{10} t_{i,N}  &=& \log_{10}(Q_{70,i,N}) -  \log_{10}(Q_{50,i,N}), \label{eq:fiterror1} \\
\Delta^- \log_{10} t_{i,N}  &=& \log_{10}(Q_{50,i,N}) -  \log_{10}(Q_{30,i,N}), \label{eq:fiterror2}
\eea

\noindent
where $Q_{30,N}$,  $Q_{50,N}$ and $Q_{70,N}$ are 30\%, 50\% and 70\% quantiles 
of the corresponding 
distribution of dissolution times $t_i$. The quantiles have been calculated with an {\sc idl}
routine by \citet{Hong2004}.
The weights used in the fitting procedure are given by

\be
w_{i,N}=1/\left[\vert \Delta^+ \log_{10} t_{i,N} \vert + \vert \Delta^- \log_{10} t_{i,N}\vert\right]^2. \label{eq:fitweights}
\ee

%As can be seen, there is an intersection
%point in Figure  \ref{fig:thnall}
%which may be used to calibrate the definition
%Eqn. (\ref{eq:t1}) of the two-body relaxation time.

The dependence of $x_{50}(\gamma, \lambda')$ on the 99\% Roche volume filling factor $\lambda'$
and the $\gamma$ factor in the Coulomb logarithm 
can be seen in Figure \ref{fig:exp}. We calculated the scaling exponents for two different 
values of the $\gamma$ parameter in the Coulomb logarithm: 
$\gamma=0.02$ \citep[multi-mass case]{Giersz1996} and
$\gamma=0.11$ \citep[equal-mass case only for comparison]{Giersz1994} 
to show the difference between
these two cases.

\newcommand{\middleincludegraphics}[2][]{%
  \raisebox{\dimexpr-0.575\height+\ht\strutbox\relax}{\includegraphics[#1]{#2}}}
  
\begin{figure*}
\begin{tabular}{cccc}
 & $t_{20}$ & $t_{50}$ & $t_{80}$ \\
$F$ &
\middleincludegraphics[angle=90,width=0.3\textwidth]{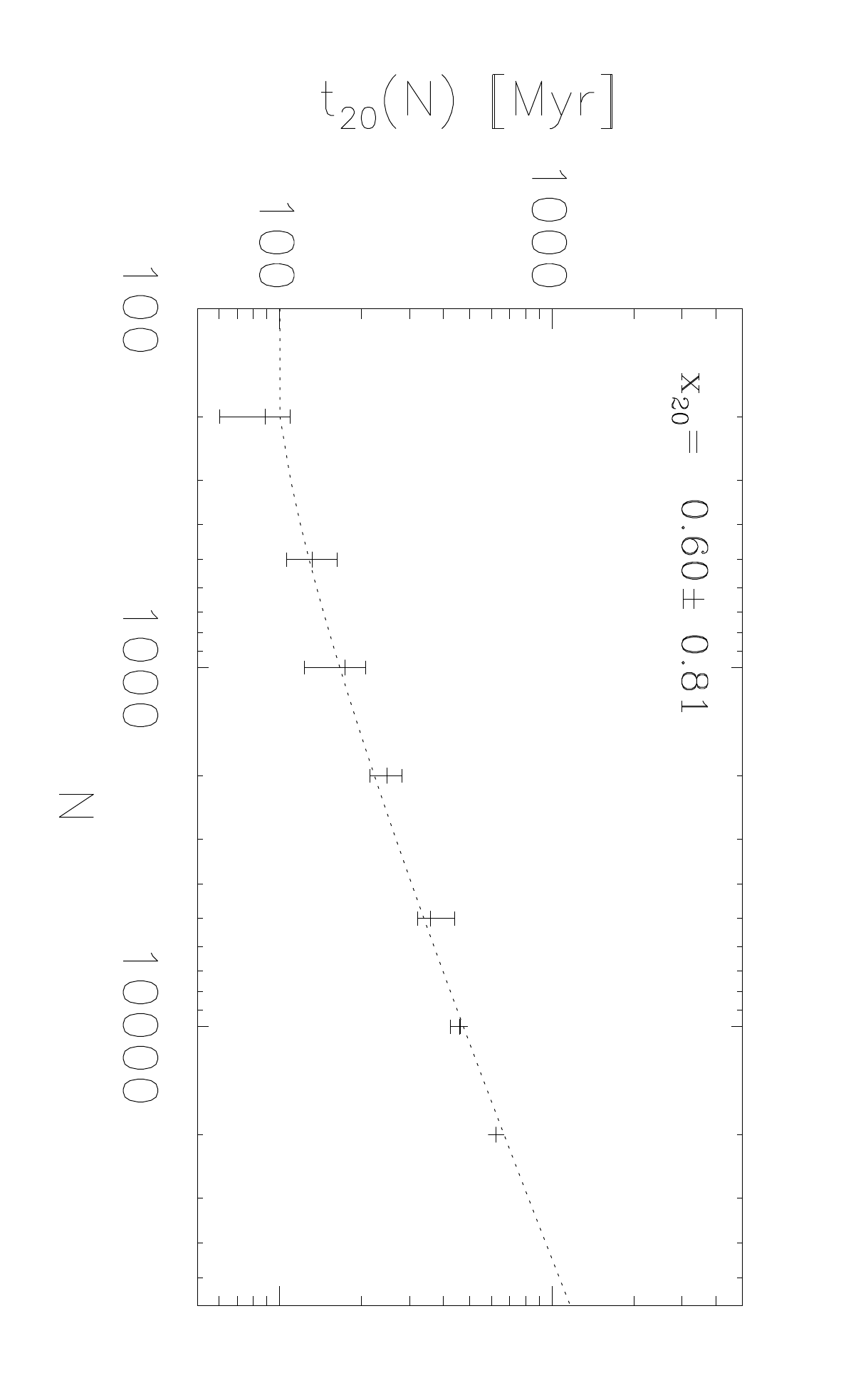} &
\middleincludegraphics[angle=90,width=0.3\textwidth]{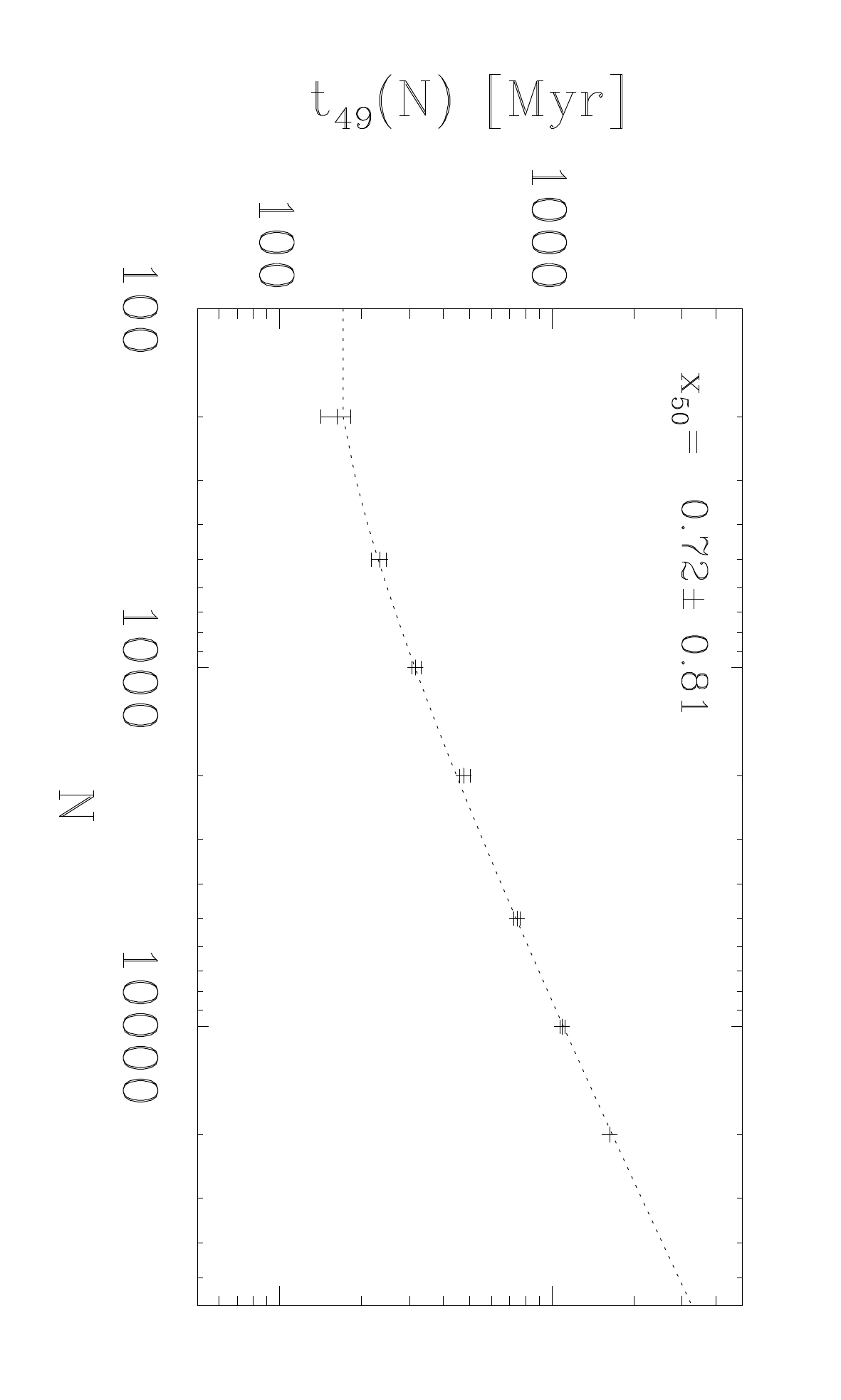} &
\middleincludegraphics[angle=90,width=0.3\textwidth]{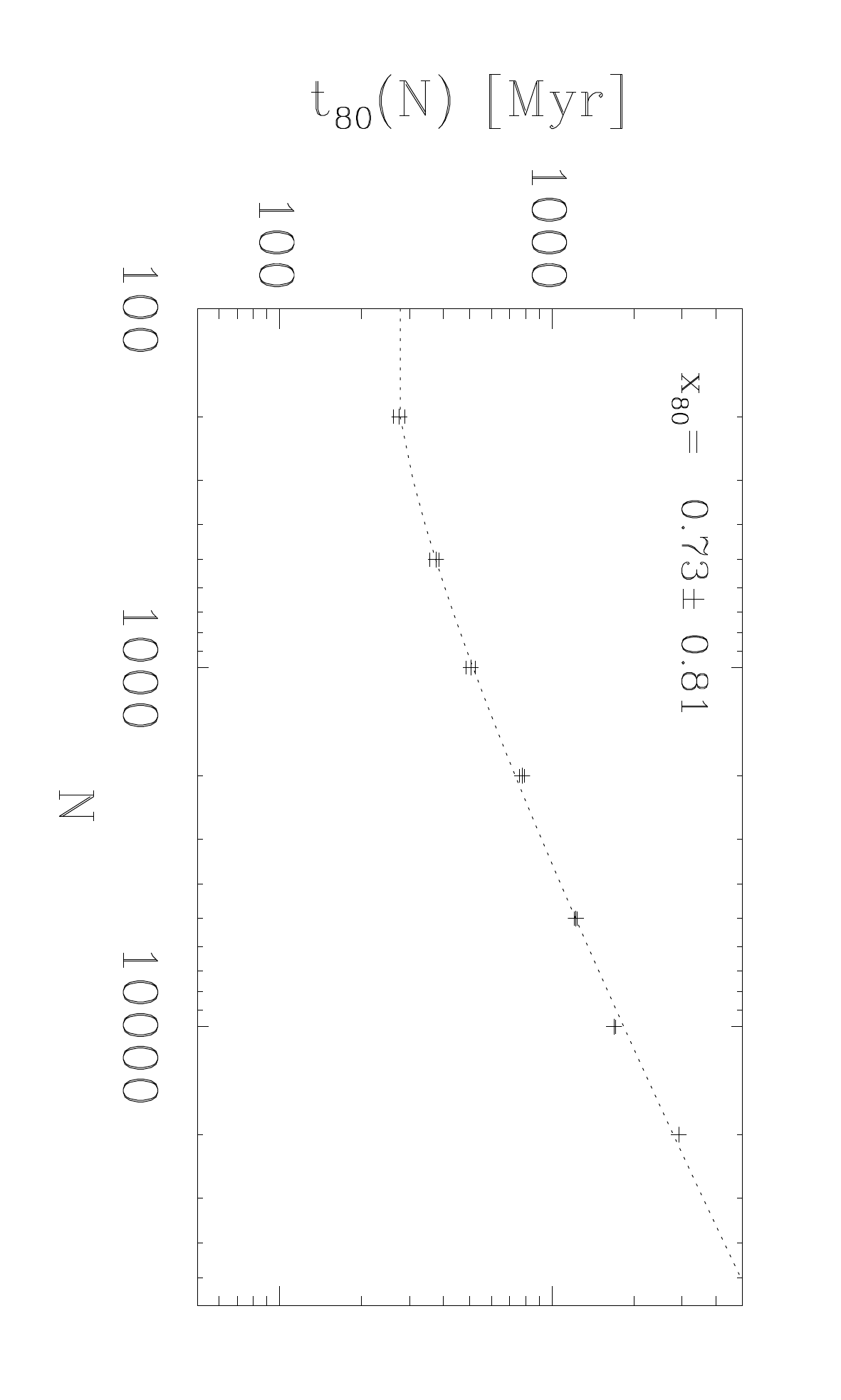} \\
$U_2$ &
\middleincludegraphics[angle=90,width=0.3\textwidth]{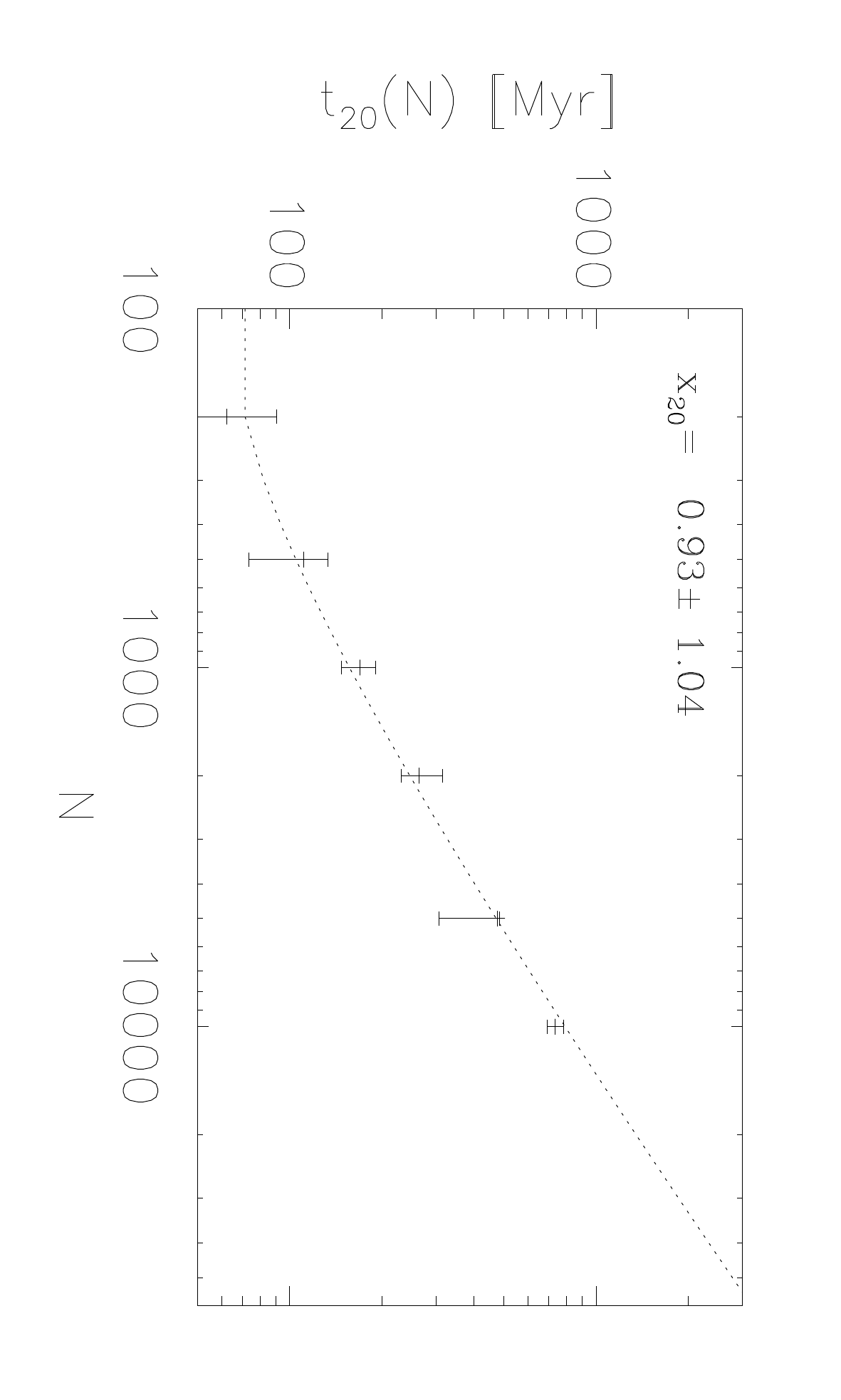} &
\middleincludegraphics[angle=90,width=0.3\textwidth]{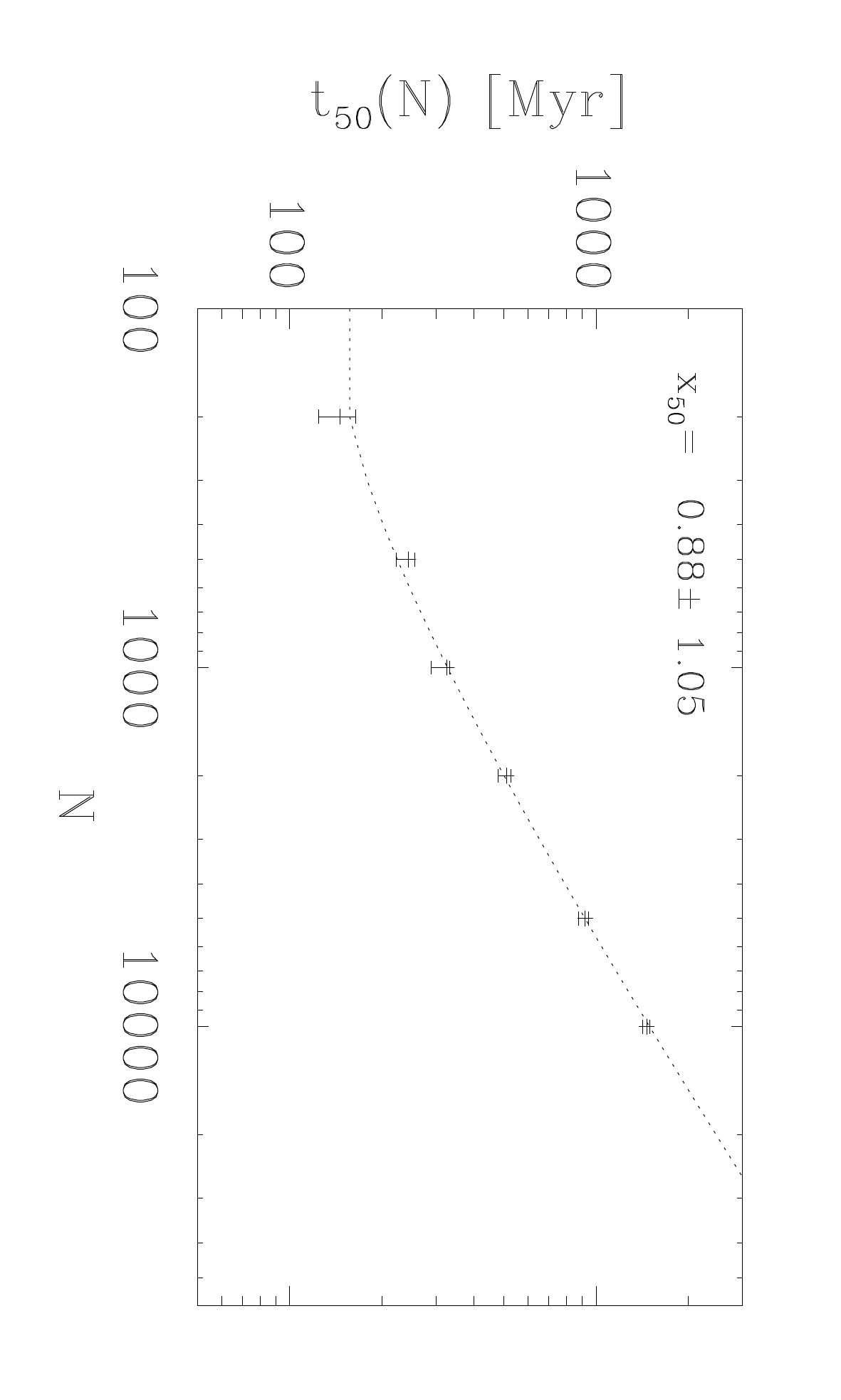} &
\middleincludegraphics[angle=90,width=0.3\textwidth]{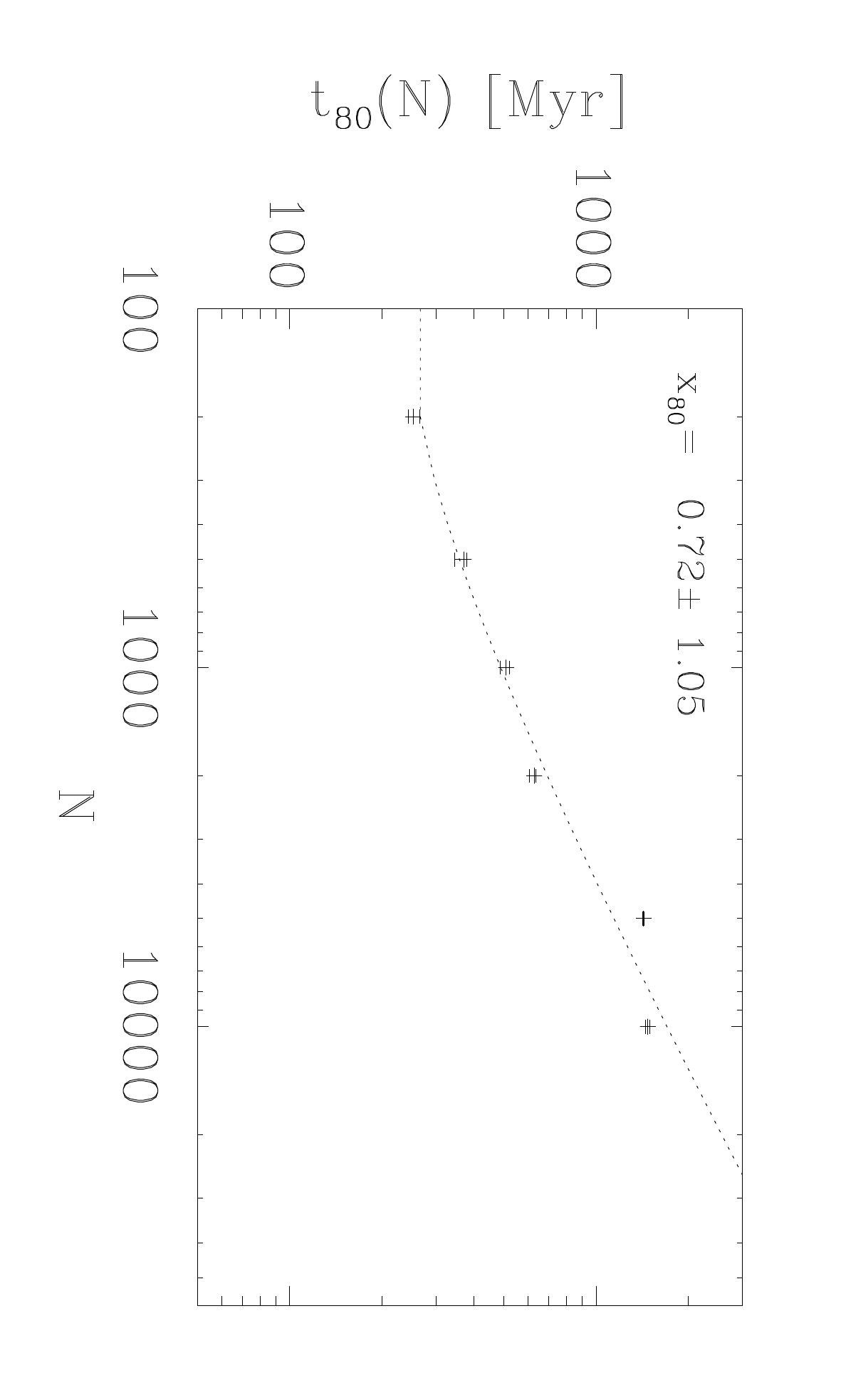} \\
$O_1$ &
\middleincludegraphics[angle=90,width=0.3\textwidth]{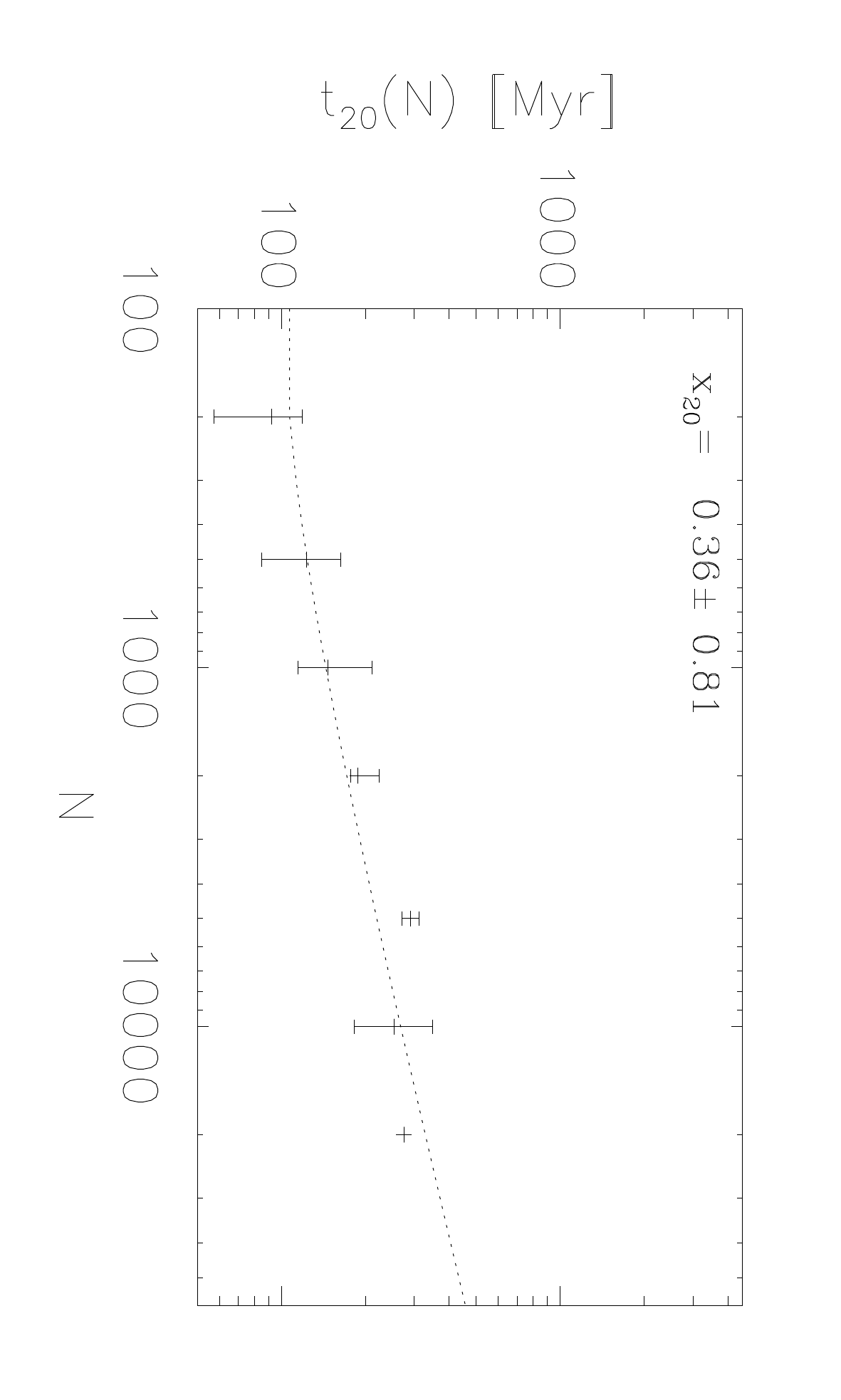} &
\middleincludegraphics[angle=90,width=0.3\textwidth]{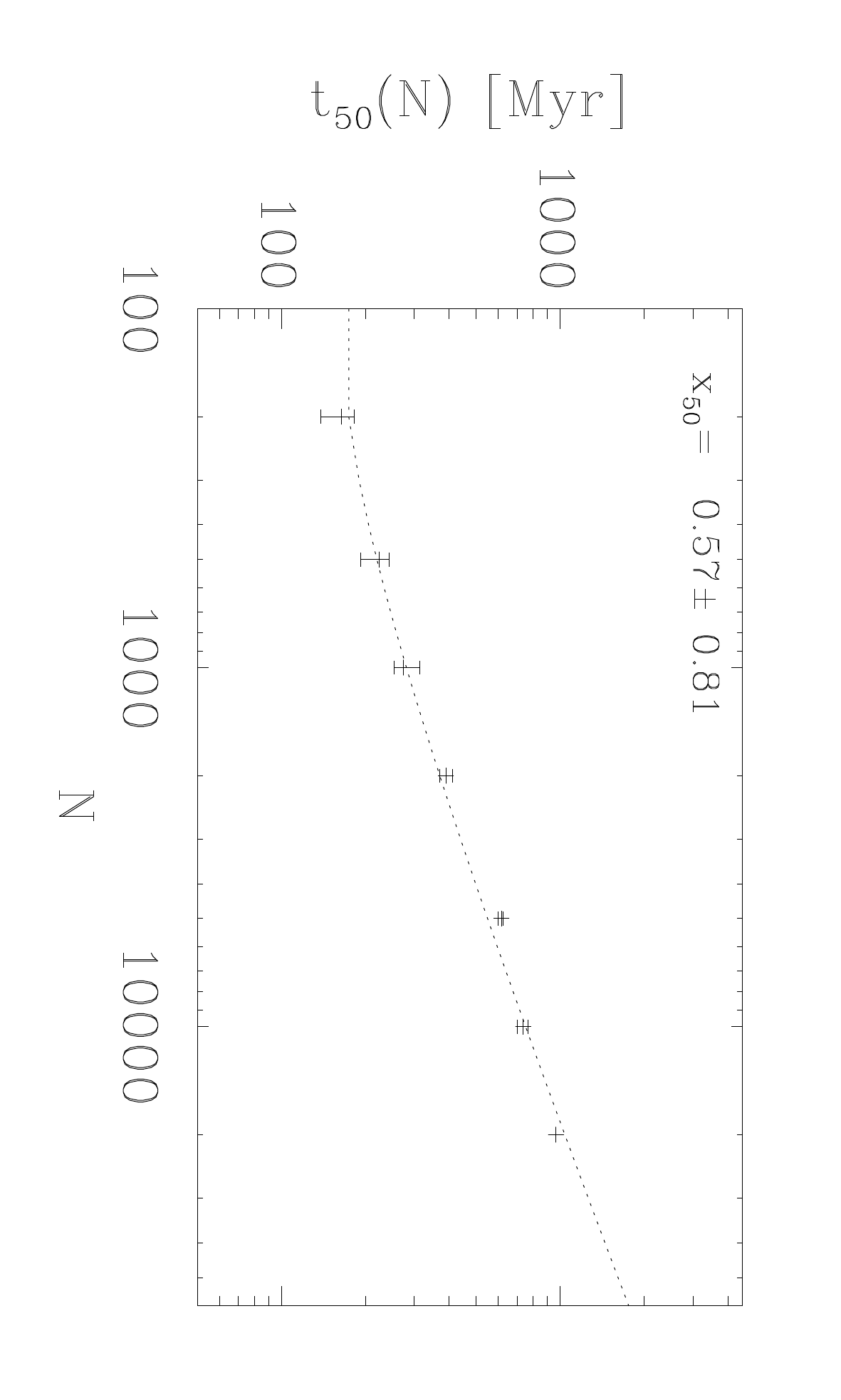} &
\middleincludegraphics[angle=90,width=0.3\textwidth]{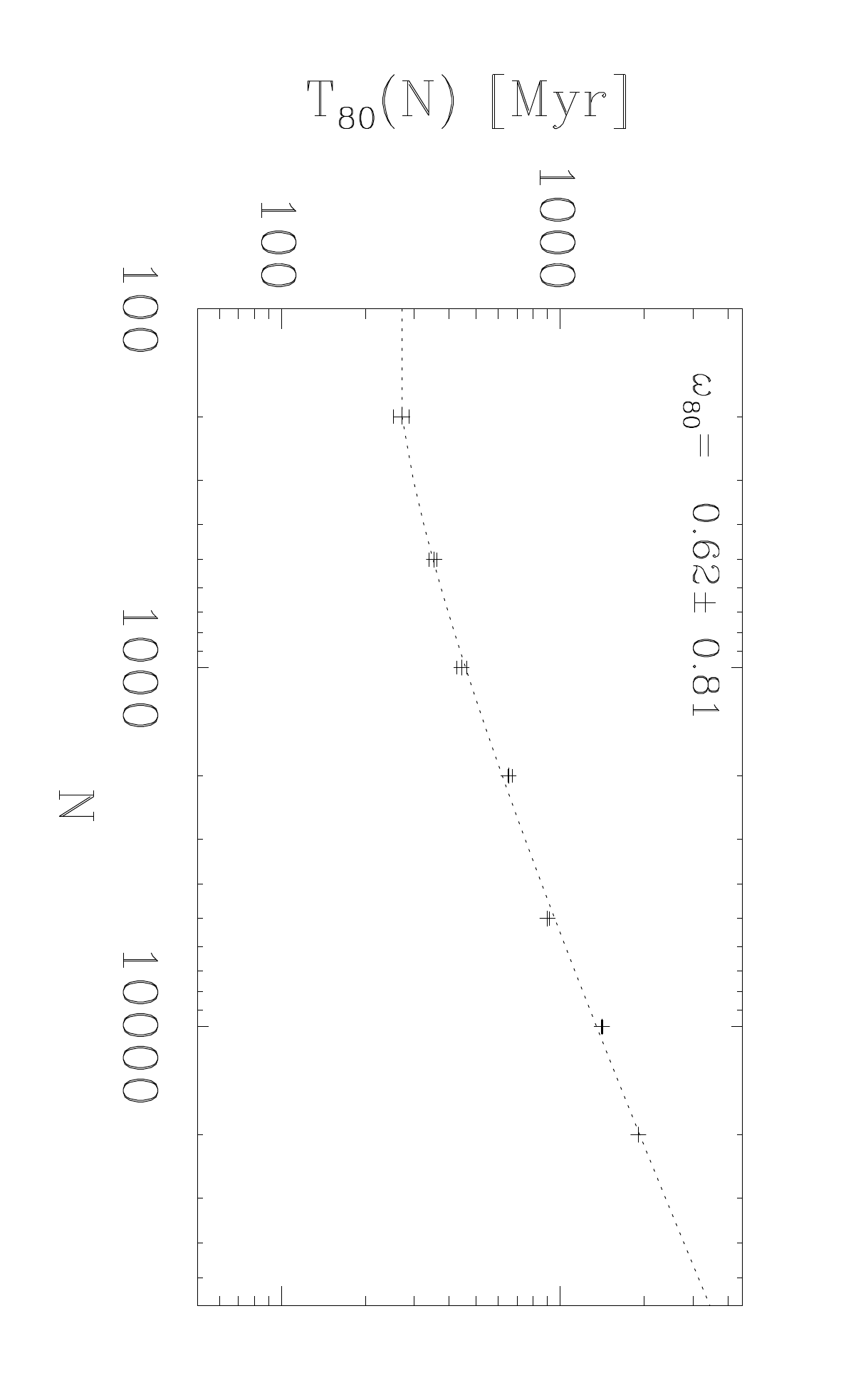} \\
$O_2$ &
\middleincludegraphics[angle=90,width=0.3\textwidth]{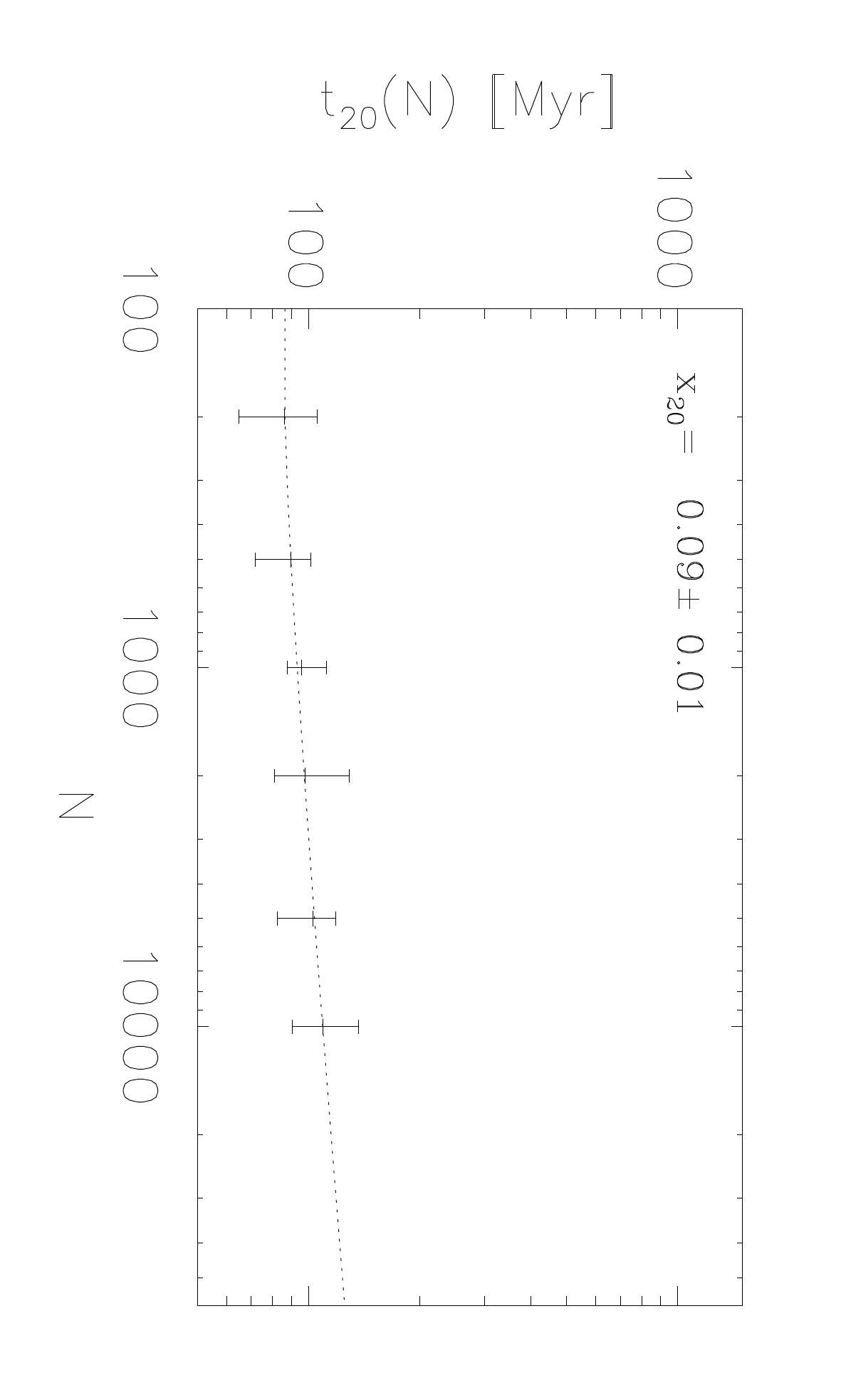} &
\middleincludegraphics[angle=90,width=0.3\textwidth]{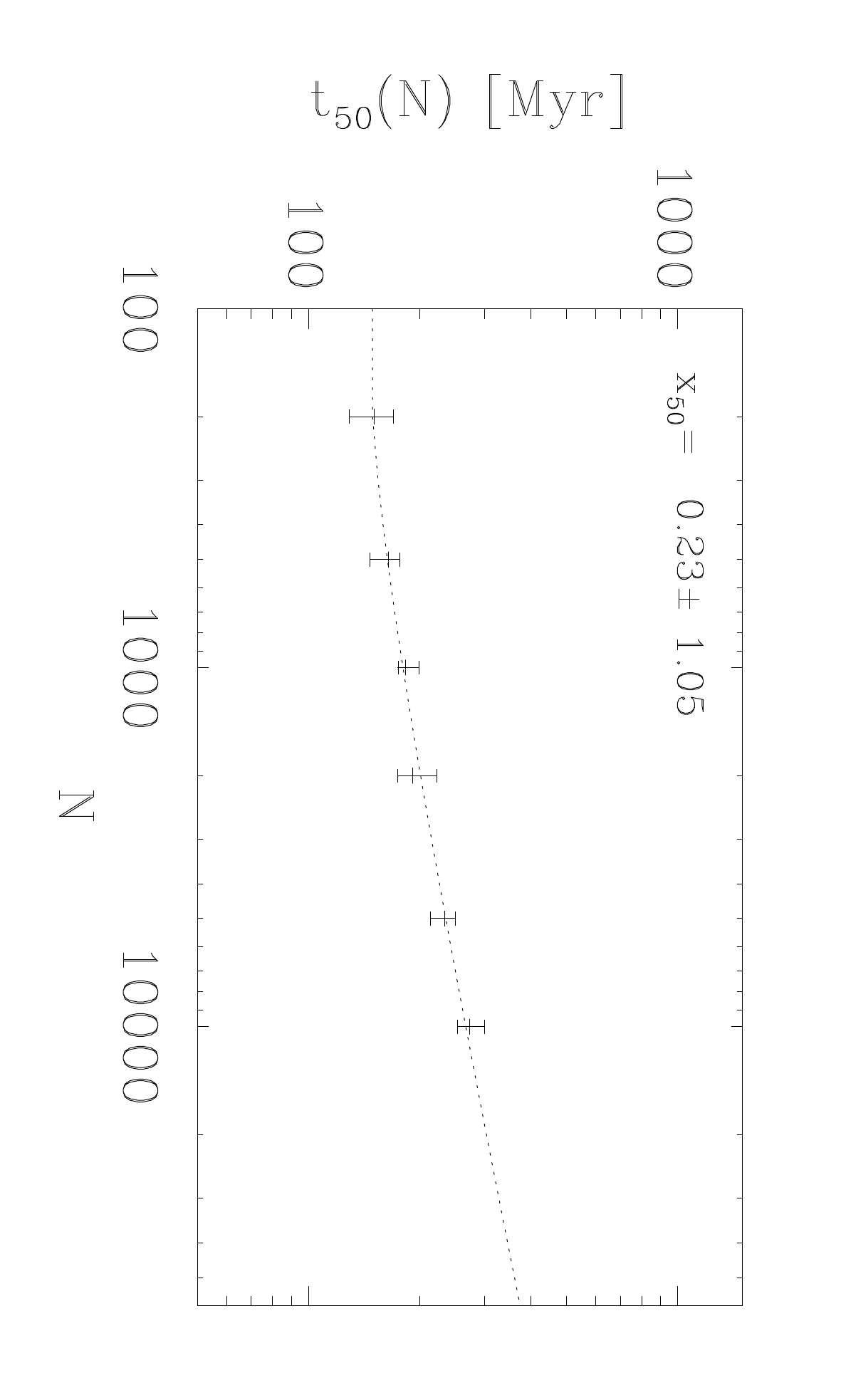} &
\middleincludegraphics[angle=90,width=0.3\textwidth]{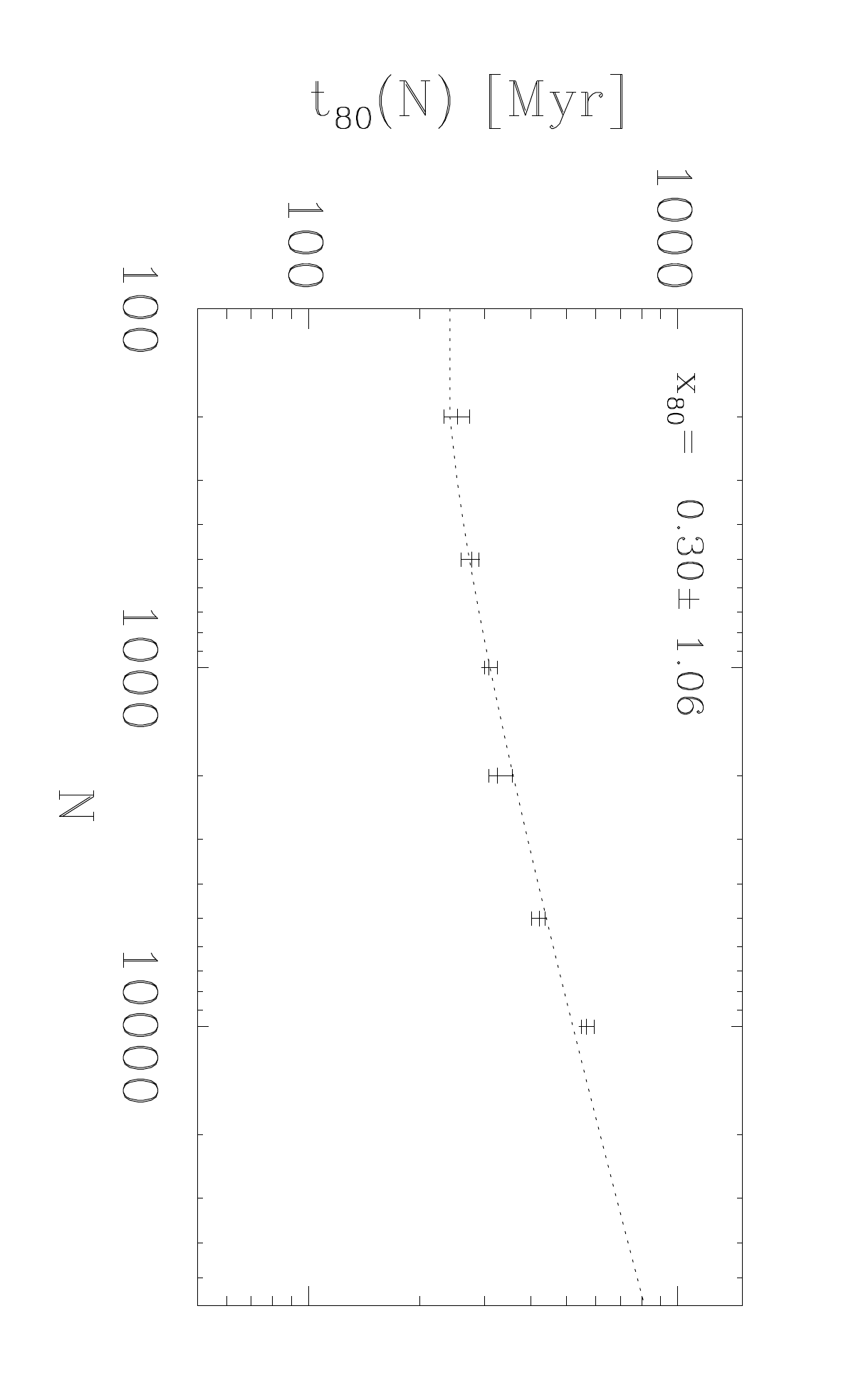} \\
\end{tabular}
\caption{Example fits.
From left to right: For $t_{20}$, $t_{50}$, $t_{80}$.
From top to bottom: For Series $F, U_2, O_1, O_2$ . The fits of $t_{20}$ can be biased due to stellar evolution mass loss. We used $\gamma=0.02$ in the Coulomb logarithm. The weights and error bars 
are calculated from the quantiles $Q_{30,N}$, $Q_{50,N}$ and $Q_{70,N}$ (see text and Eqns. (\ref{eq:fiterror1}) - (\ref{eq:fitweights})).} 
\label{fig:tifitex205080}
\end{figure*}

\begin{figure*}
\includegraphics[angle=90,width=1.0\textwidth]{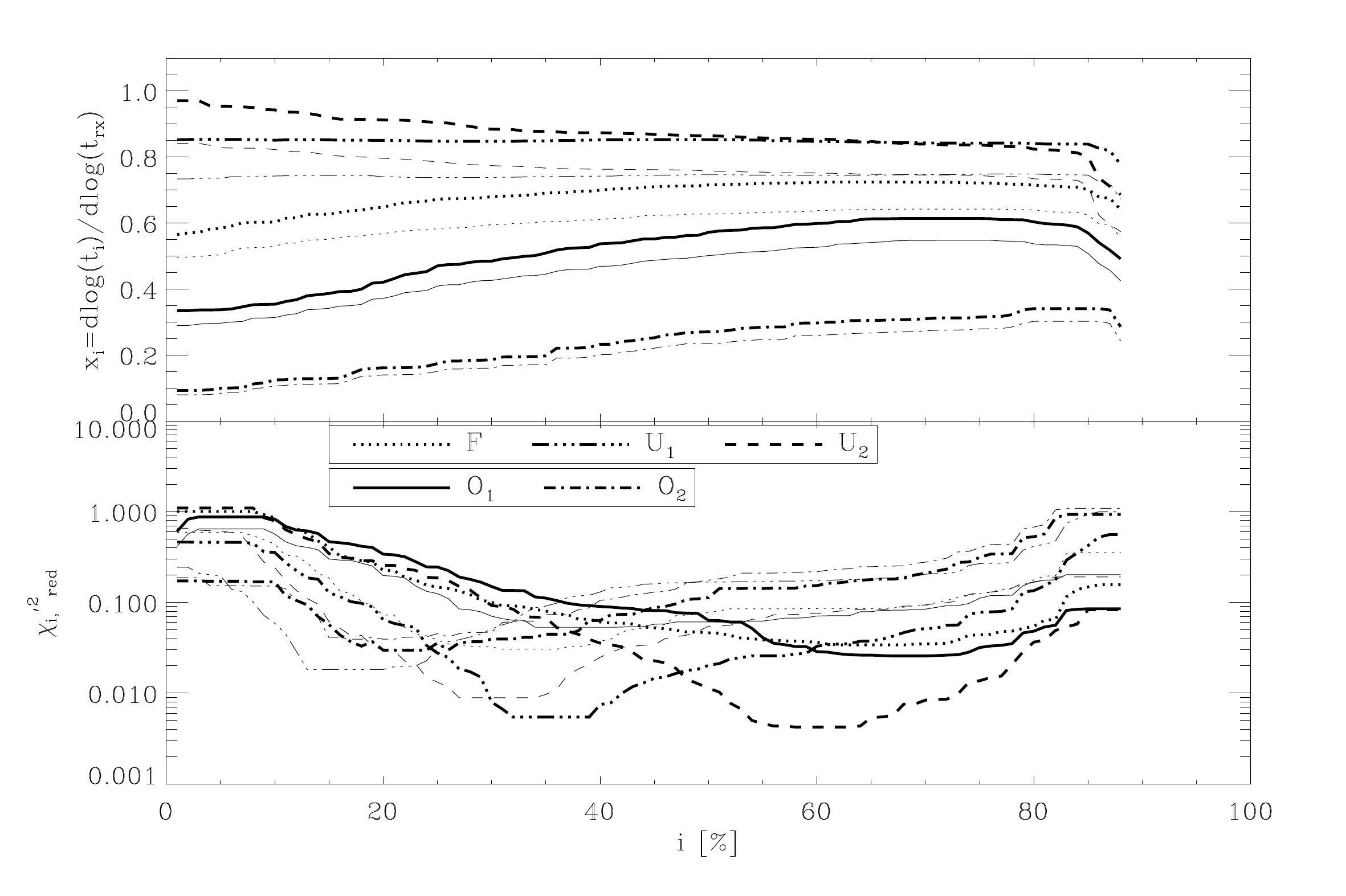} 
\caption{Top panel: Scaling exponents $x_i=d\log(t_{\rm i})/d\log(t_{\rm rh})$.
The time $t_i$ is defined as the time when the cluster has lost $i\%$ of its initial particle number.  
We used a median smoothing with a smoothing width of 11 in i.  
Bottom panel: The corresponding $\chi'^2_{red,i}$ values. For low i we find large  $\chi'^2_{red,i}$ due to the stellar evolution mass loss.  The half-number time seems to be a robust
measure with respect to the $\chi'^2_{red,i}$ values. Thick lines: $\gamma=0.02$; thin lines: $\gamma=0.11$.
The weights are calculated from the quantiles $Q_{30,N}$, $Q_{50,N}$ and $Q_{70,N}$ (see text and Eqn. (\ref{eq:fitweights})).} 
\label{fig:xiallq2}
\end{figure*}

Figure \ref{fig:tifitex205080} shows example fits for the times $t_{20}$, $t_{50}$ (half-number time)
and $t_{80}$ for Series $F, U_2, O_1, O_2$. Here we used only $\gamma=0.02$ in the Coulomb
logarithm. For the least-squares-fitting, we used the {\sc mpfit} package in
{\sc idl} \citep[for the Levenberg-Marquardt algorithm]{Markwardt2009, More1978}. 

Figure \ref{fig:xiallq2} shows all scaling exponents

\be
x_i=\frac{d\log(t_i)}{d\log(t_{\rm rx})}
\ee 

\noindent
for $i=1-90$ (in percent), together with the corresponding $\chi^2$ values. The time $t_{i}$ is defined as the time when the cluster has lost $i\%$ of its initial particle number.
We calculated the scaling exponents for two different values of the 
$\gamma$ parameter in the Coulomb logarithm. We used

\be
\chi'^2_{red,i} = \frac{1}{N_{\rm dof}}\sum_{k=1}^{N_{\rm dof}}( (y_{i,k}-f_{i,k})^2 \times \vert w_{i,k} \vert )
\ee

\noindent
with the weights of Eqn. (\ref{eq:fitweights}), where 
$f_{i,k}$ is the value of the fitted power law function Eqn. (\ref{eq:ansatz}), 
the $y_{i,k}$ are the data and $N_{\rm dof}$ is the number of degrees of freedom \citep{Markwardt2009}.
%%AJ
From Figure \ref{fig:xiallq2} we can see that the power law index $x_i$ is a weak function of the mass loss fraction. The $\chi'^2_{red,i}$ is not easy to interpret since the quantiles of the 
corresponding statistic with the weights of Eqn. (\ref{eq:fitweights}) are not known. 
However, one can gain an insight regarding the
relative behaviour of $\chi'^2_{red,i}$ for different $t_{\rm i}$'s. 
It can be seen that $t_{\rm 50}$ is a robust measure for the dissolution time.
%Additionally, the value of $\gamma=0.02$, which should correctly describe the impact of 2-body %relaxation for clusters with a stellar mass spectrum, leads to a cleaner power law dependence in %the later evolution phase.
%For a large range between $t_{\rm 20}$ and  $t_{\rm 80}$ the power law fit has a similar %quality.
%%AJ end

\begin{figure}
\includegraphics[angle=90,width=0.45\textwidth,height=0.9\textheight]{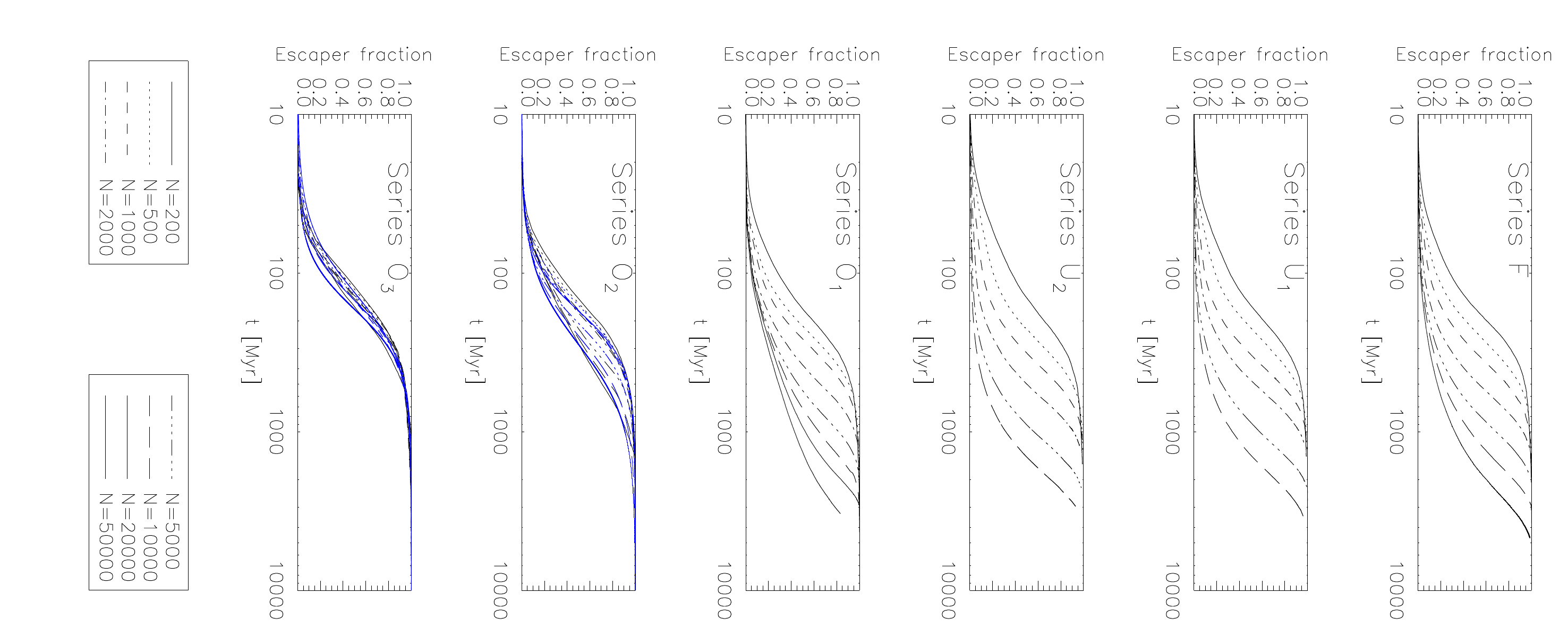} 
\caption{
Evolution of the escaper fraction. From left to right in each panel: 
$N=200, 500, 1k, 2k, 5k, 10k, 
20k, 50k$ and fits with Eqn. (\ref{eq:psls}). The time axis has been switched from linear to logarithmic.
The log-logistic growth occurs in the  Roche volume overfilling limit. The corresponding
fits are shown (blue lines).}
\label{fig:rlrjpslog}
\end{figure}

\subsection{Log-logistic growth}

\label{sec:pslg}

\begin{table}
\caption{Averaged parameters $(\log_{10}(t_{50} \ [\mathrm{Myr}]), \kappa)$ from 
fits with Equation (\ref{eq:psls2}).}
\label{tab:psl}
\begin{center}
\hspace{-1.5cm}
\begin{tabular}{ccccccccc}
\hline
Series & $\log_{10}(t_{50} \ [\mathrm{Myr}])$ & $\kappa$ \\
%Series / $N$ & 200 & 500 & 1k & 2k & 5k & 10k & 20k \\
%\ \ \ \ \ \ \ \ \ \ \ / $n$ & 128 & 64 & 32 & 16 & 8 & 4 & 2 & 1 \\
\hline
$O_2$ & $2.30 \pm 0.09$ &  $2.25 \pm 0.28$ \\
$O_3$ & $2.16 \pm 0.04$ &  $2.18 \pm 0.20$ \\
%D & (2.15,11.5) & (2.18,12.1) & (2.22,12.5) & (2.23,13.0) & (2.33,13.1) & (2.39,12.3) & (2.44,11.9)  \\
%E & (2.09,10.1) & (2.10,11.0) & (2.14,10.8) & (2.14,12.0) & (2.11,10.4) & (2.15,12.7) & (2.21,12.2) \\
\hline
\end{tabular}
\end{center}
\end{table}

Figure \ref{fig:rlrjpslog} shows the time evolution of the
escaper fraction $N_e(t)/N_0= 1-N(t)/N_0$ with 
a logarithmic time axis, where $N(t)$ (and $N_0$) are taken to be the current (initial) 
particle number within three times the Jacobi radius.

In the Roche volume overfilling limit the evolution of the escaper fraction can,
at least empirically, be 
approximately described by a log-logistic differential equation in logarithmic time,

\be
\frac{d\ln N_e}{d\ln t} = \kappa \left(1-\frac{N_e}{N_0}\right) = \kappa \frac{N}{N_0}. \label{eq:psl}
\ee

\noindent
The solution is given by

\be
\frac{N_e(t)}{N_0} = \frac{t^\kappa}{t_{50}^\kappa + t^\kappa}.\label{eq:psls}
\ee

\noindent
The evolution of $N(t)/N_0$ can then be described by
the law 

\be
\frac{N(t)}{N_0} =  \frac{t_{50}^\kappa}{t_{50}^\kappa + t^\kappa}.\label{eq:psls2}
\ee

\noindent
The best-fit exponent $\kappa$ and the best-fit half-number time $t_{50}$ 
can be determined for the parameter space covered in this study. 
Table \ref{tab:psl} shows the parameters of least-quares fits with the Equation

\be
\log t = \log t_{50} + \frac{1}{\kappa}\log\left(\frac{N_e/N_0}{1-N_e/N_0}\right).
\ee

\noindent
For the least-squares-fitting, we used the {\sc mpfit} package in
{\sc idl} \citep[for the Levenberg-Marquardt algorithm]{Markwardt2009, More1978}. 
The fits are shown in the two lowest panels on the right-hand side of Figure \ref{fig:rlrjpslog}
(for series $O_2$ and $O_3$). For the series shown in the upper panels we suspect a transition
from log-logistic to logistic with some other contribution from left to right, 
where the factor $1/t$ is due to the stellar evolution \citep{Lamers2010} 
and the other contribution is due to the relaxation-driven evolution.

%Begin AJ slightly changed by AE
There are two competing processes for populating the potential escaper reservoir
above the critical Jacobi energy. Firstly, scattering by stellar encounters
scales with the relaxation time and depends on the particle number $N$.
Secondly, cluster mass loss by stellar evolution and by escaping stars
lowers the cluster potential well and lifts the critical Jacobi energy,
which shifts new stars above the critical value. The fractional mass loss
rate is independent of $N$ and results in a particle loss rate proportional to
$N$, to the number of escapers $N_e$ and via stellar evolution to $1/t$
\citep{Lamers2010}. For the overfilling clusters (series $O_2$, $O_3$) 
two-body encounters are negligible leading to the log-logistic behaviour.
For the more concentrated clusters, where two-body encounters play an important role, the interplay of the different timescales is discussed in detail in \citet{Lamers2010}.
Only for very large $N$, where the relaxation time is very long,
the factor $1/t$ by stellar evolution in the dissolution timescale becomes relevant again.
This leads to a reduction of $t_{50}$ with respect to the power law dependence derived in Eqn. (6) for small filling factors.
%End AJ slightly changed by AE

%It is interesting to note that Eqn. (\ref{eq:psl}) has two fixed points at $N_e/N_0 = 0$ and $N_e/N_0 = 1$.
%The Jacobi matrix element $J_{00}$ is equal to the eigenvalue

%\be
%\eta = J_{00} = \kappa\frac{t_0}{t}\left(1-2\frac{N_e}{N_0}\right).
%\ee

%\noindent
%For $\kappa t_0/t>0$ we find that the fixed point $N_e/N_0 = 0$ is unstable while the fixed point
%$N_e/N_0 = 1$ is stable.  This is in harmony with the fact that Roche volume overfilling star clusters 
%are unstable with respect to dissolution.

\section{Conclusions}

We have carried out a parameter study of open star 
clusters with the parameters $(N,\lambda= r_{t}/r_J)$.
We have found the following results:

\begin{enumerate}

\item The $N$-dependence of the dissolution time in units of the two-body relaxation time
is well fitted by a power law.  The power law index is a function of the Roche volume filling factor $\lambda$
and the $\gamma$ factor in the Coulomb logarithm. 
It decreases with increasing  $\lambda$ with the limiting value of zero in the overfilling limit.
Particularly in the underfilling limit, the power law index has been found to depend 
on the value of the $\gamma$ parameter adopted in the Coulomb logarithm.

\item Our study suggests that open star clusters in the Roche volume overfilling regime dissolve mainly due to the changing cluster potential
%%AJ
by lifting stars above the decreasing critical Jacobi energy.
%%AJ end
We call this mechanism ``mass-loss driven dissolution'' 
in contrast to the ``two-body relaxation driven dissolution'' which occurs from the 
Roche volume underfilling regime up to the 
Roche volume filling case \citep[see also][based on simpler models]{Whitehead2013}. 

\item In the Roche volume overfilling limit the escaper fraction $N_e(t)/N_0$ 
obeys, at least empirically, approximately a log-logistic differential equation in logarithmic time.

\end{enumerate}

\noindent
We make the following remarks: 
\begin{enumerate}

\item The mass-loss driven dissolution provides a mechanism, which is responsible 
for the dissolution of OCs which have survived the gas expulsion phase with a relatively large initial half-mass radius as observed for example in the Pleiades cluster \citep{Converse2010}. 
It is a viable mechanism besides the dissolution due to encounters with giant molecular clouds 
\citep{Wielen1971, Wielen1985}. 

\item The mass-loss driven dissolution in combination with a
large scatter in Roche volume filling factors \citep{Ernst2013} also naturally explains 
the large scatter in the lifetimes of open clusters \citep{Wielen1971}. In a future paper we plan to investigate the impact of the newly found $N$-independence of the dissolution timescale for the overfilling clusters on the CFR using the observed mass-age distribution of OCs in the solar neighbourhood.

\item Due to the intricacy of the problem 
a detailed theory of mass-loss driven dissolution, which explains the 
transition from the over- to underfilling limits in terms of the dependence of $x_{50}$ on $\lambda$ as shown in Figure \ref{fig:exp} has not yet been developed and is beyond the scope
of the present experimental study.  

\end{enumerate}

\acknowledgements

The main set of simulations and data analysis was performed on the
GPU accelerated supercomputers {\tt titan},
{\tt hydra} and {\tt kepler} of the GRACE project led by 
Prof. Dr. Rainer Spurzem, funded under the grants 
I/80041-043, I/84678/84680 and I/81 396 of the Volkswagen foundation 
and 823.219-439/30 and /36 of the Ministry of Science, 
Research and the Arts of Baden-W\"urttemberg). 
A few standard desktop PCs in the first author's office were also used.

PB acknowledges the support by Chinese Academy of Sciences
through the Silk Road Project at NAOC and through the Chinese 
Academy of Sciences Visiting Professorship program for Senior 
International Scientists.  PB also acknowledges the special support by the 
NAS Ukraine under the Main Astronomical Observatory GPU/GRID computing 
cluster project.  

AE acknowledges support by grant JU 404/3-1 of the Deutsche 
Forschungsgemeinschaft (DFG) and would like to thank 
Dr. Holger Baumgardt for two discussions.

PB and AE further acknowledge the financial support by the Deutsche 
Forschungsgemeinschaft (DFG) through Collaborative 
Research Center (SFB 881) "The Milky Way System" 
(subprojects B1 and Z2) at the Ruprecht-Karls-Universit\"at 
Heidelberg.

\appendix

\section{Resonance condition}

\label{sec:rescond}

It is possible to write down a resonance condition \citep{Ernst2013},

\bea
\frac{m}{n}&=&\frac{\Omega}{\omega_{\rm res}} = \left( \frac{4-\beta_C^2}{2} \right)^{1/2}  \left(\frac{r_{\rm res}}{r_J}\right)^{3/2}, \label{eq:rescond}
%&=&\frac{2}{\pi^2} \left( \frac{4-\beta_C^2}{2} \right)^{1/2}  \left(-\frac{3GM_{\rm cl}}{2E_{\rm J, crit} r}
%\right)^{3/2} 
\eea

\noindent
where $m$ and $n$ are natural numbers and 
the orbital frequencies $\omega_{\rm res} = \omega(r_{\rm res})$ and $\Omega=2\pi/t_{\rm orb}$ 
are related to the orbital time of a star in the star cluster and the 
orbital time of the  star cluster on a circular orbit
around the galaxy, respectively.
$\beta_C$ is the ratio between epicyclic and circular frequency and $r_{\rm res}$ is a resonance
radius. For a flat rotation curve we have $\beta_C=\sqrt{2}$. For the Milky Way model with 
the parameters given in Table \ref{tab:gal-par} we have $\beta_C\approx 1.37$ at $R_g=8$ kpc.
Therefore at certain values of $\lambda=r_h/r_J$ resonance effects may occur
and play a role in the evolution. 

The central periodic orbit in the largest regular (i.e. non-chaotic)
island in the Poincar\'e surfaces of section at the 
critical Jacobi energy has $(r_{\rm res}/r_J) = 0.345$. This island corresponds to
quasiperiodic retrograde orbits \citep{Fukushige2000, Ernst2008}. 
It may be that the corresponding resonance is linked with the 
properties of star clusters and connected with the occurence of two discrete 
types of star clusters, open and globular clusters. Moreover, the location of the resonance may be 
used to  calibrate the relation (\ref{eq:rescond}) more precisely for star clusters, i.e.
to determine $r_{\rm res}$ as a function of $r_h$.

\section{Stability curve}

\label{sec:stabrel}

\begin{figure}
\includegraphics[angle=90,width=0.5\textwidth]{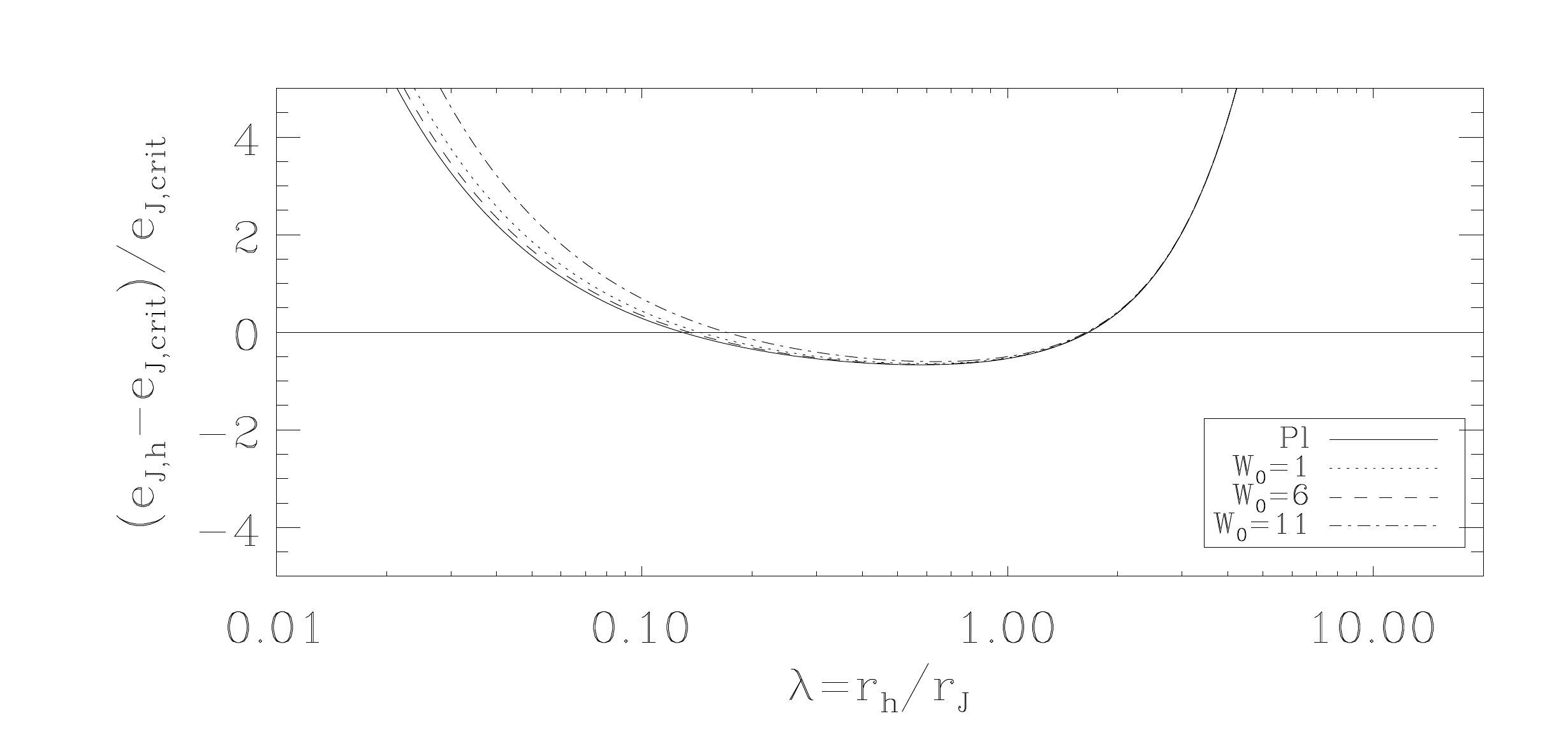} 
\caption{Stability curve for a Plummer model and King models. Value of $r_h/r_V$ for $W_0=11$ 
by G\"urkan and Freitag, priv. comm.} 
\label{fig:dej}
\end{figure}

In connection with the Roche volume filling factor a stability curve can be 
derived for star clusters. Such a curve has been derived for the first time in 
\citet[][their Eqn. 28; see also their Figure 13]{Fukushige1995} and may
therefore be called the ``Fukushige-Heggie stability curve'' for star clusters.
The Jacobi energy of a typical star at the half-mass radius is given by

\be
e_{J, h} = \frac{K}{M_{\rm cl}} + \frac{W}{M_{\rm cl}} - \frac{1}{2} \frac{GM}{r_h}\lambda^3 \\
\ee

\noindent
with $\lambda=r_h/r_J$,
where $K$ and $W$ are the kinetic and potential energies, respectively.
The critical Jacobi energy is given by

\be
e_{\rm J, crit} = -\frac{3}{2}\frac{G M_{\rm cl}}{r_J}. \label{eq:ejcrit}
\ee

\noindent
We obtain for a Plummer model with

\be
K=-\frac{W}{2}=\frac{3\pi}{64}\frac{GM_{\rm cl}^2}{r_{\rm Pl}}, \label{eq:virrel}  
\ee
\noindent
and $r_h/r_{\rm Pl}\approx 1.305$ the stability curve

\be
\frac{e_{J, h}-e_{J,\rm crit}}{e_{J,\rm crit}} \approx \frac{1.305\cdot\pi}{32 \lambda} 
+ \frac{\lambda^2}{3} - 1 \label{eq:plumstab}
\ee

\noindent
For a King model with $W_0=6$ we have

\be
K = -E = \frac{GM_{\rm cl}^2}{4 r_V} \label{eq:virrel2} 
\ee

\noindent
with $r_h/r_V\approx 0.804$ \citep[Half-mass radius in $N$-body units $G=M_{\rm cl}=-4E=1$,][Table 1]{Guerkan2004}, where $E$ is the total energy. We obtain 
the stability curve

\be
\frac{e_{J, h}-e_{J,\rm crit}}{e_{J,\rm crit}} \approx \frac{0.804}{6 \lambda} 
+ \frac{\lambda^2}{3} - 1 \label{eq:kingstab}
\ee

\noindent
The stability curves in Eqns. (\ref{eq:plumstab}) and (\ref{eq:kingstab}) are
shown in Figure \ref{fig:dej} for a Plummer model and 3 King models and
may be also connected with the occurence of two discrete 
types of star clusters, open and globular clusters. The value of this function measures the relative
Jacobi energy difference between a typical star at the half-mass radius and the 
critical Jacobi energy. Note that  Eqn. (\ref{eq:ejcrit}) and the virial relations in Eqns. (\ref{eq:virrel}) and (\ref{eq:virrel2}) break down for large $\lambda$,
i.e. only the first zero at small $\lambda$ may be of physical significance.
We expect relaxation driven
dissolution in the regime to the left of the first zero and mass-loss driven dissolution in the
regime to the right of the first zero.

\label{lastpage}


\begin{thebibliography}{99}
%\hypertarget{mybib}{}
%\pdfbookmark[0]{Bibliography}{mybib} 
\bibitem[\protect\citeauthoryear{Aarseth}{2003}]{Aarseth2003}
Aarseth S. J. 2003, {\it Gravitational $N$-body simulations -- 
Tools and Algorithms}, Cambridge Univ. Press, Cambridge, UK
\bibitem[\protect\citeauthoryear{Ahmad \& Cohen}{1973}]{Ahmad1973}
Ahmad A., Cohen L., 1973, J. Comp. Phys., 12, 389
\bibitem[\protect\citeauthoryear{Baumgardt}{2001}]{Baumgardt2001}
Baumgardt H, 2001, MNRAS, 325, 1323
\bibitem[\protect\citeauthoryear{Baumgardt \& Makino}{2003}]{Baumgardt2003}
 Baumgardt H., Makino J., 2003, MNRAS, 340, 227
\bibitem[\protect\citeauthoryear{Binney \& Tremaine}{1987}]{Binney1987}
Binney J., Tremaine S., 1987, {\it Galactic dynamics}, Princeton Univ. Press, USA
\bibitem[\protect\citeauthoryear{Binney \& Tremaine}{2008}]{Binney2008}
Binney J., Tremaine S., 2008, {\it Galactic dynamics}, second edition, Princeton Univ. Press, USA
\bibitem[\protect\citeauthoryear{Boutloukos \& Lamers}{2003}]{Boutloukos2003}
Boutloukos, S. G.; Lamers, H. J. G. L. M., 2003, MNRAS 336, 1069
\bibitem[\protect\citeauthoryear{Chandrasekhar}{1943}]{Chandrasekhar1943}
 Chandrasekhar S., Ap. J. 97, 255 (1943).
\bibitem[\protect\citeauthoryear{Converse \& Stahler}{2010}]{Converse2010}
Converse, J.M., Stahler, S.W., 2010, MNRAS, 405, 666
\bibitem[\protect\citeauthoryear{Engle}{1999}]{Engle1999}
 Engle K. A., 1999, PhD thesis, Drexel University
\bibitem[\protect\citeauthoryear{see also Ernst et al.}{2008}]{Ernst2008}
Ernst A., Just A., Spurzem R., Porth O., 2008, MNRAS 383, 897
\bibitem[\protect\citeauthoryear{Ernst, Just \& Spurzem}{2009}]{Ernst2009a}
Ernst A., Just A., Spurzem R., 2009, MNRAS, 399, 141
\bibitem[\protect\citeauthoryear{Ernst}{2009}]{Ernst2009b}
Ernst A., 2009, PhD thesis, University of Heidelberg, Germany
\bibitem[\protect\citeauthoryear{Ernst et al.}{2010}]{Ernst2010}
Ernst A., Just A., Berczik P., Petrov M. I., 2010, A\&A, 524, A62
\bibitem[\protect\citeauthoryear{Ernst et al.}{2011}]{Ernst2011}
Ernst A., Just A., Berczik P., Olczak C., 2011, A\&A, 536, A64
\bibitem[\protect\citeauthoryear{Ernst \& Just}{2013}]{Ernst2013}
Ernst A., Just A., 2013, MNRAS, 429, 2953
\bibitem[\protect\citeauthoryear{Fanning}{2011}]{Fanning2011}
Fanning D. W., 2011, Coyote's Guide To Traditional IDL Graphics, Coyote Book Publishing
\bibitem[\protect\citeauthoryear{Fellhauer et al.}{2003}]{Fellhauer2003}
Fellhauer M., Lin D. N. C., Bolte M., Aarseth S. J., Williams K. A., 2003, Ap. J., 595, 53
\bibitem[\protect\citeauthoryear{Heggie \& Mathieu}{1986}]{Heggie1986} 
Heggie D. C., Mathieu R. D., 1986, in Hut P., McMillan S., eds., LNP 267,
{\it The Use of Supercomputers in Stellar Dynamics Standardised Units and
Time Scales}, Springer Verlag, Berlin, p. 233
\bibitem[\protect\citeauthoryear{Fellhauer \& Heggie}{2005}]{Fellhauer2005}
Fellhauer, M., Heggie, D.C., 2005, A\&A 435, 875
\bibitem[\protect\citeauthoryear{Fukushige \& Heggie}{1995}]{Fukushige1995}
Fukushige T., Heggie D. C., 1995, MNRAS 276, 206
\bibitem[\protect\citeauthoryear{Fukushige \& Heggie}{2000}]{Fukushige2000}
Fukushige T., Heggie D. C., 2000, MNRAS, 318, 753 
\bibitem[\protect\citeauthoryear{Gaburov, Harfst \& Portegies Zwart}{2009}]{Gaburov2009}
Gaburov, E., Harfst S., Portegies Zwart S., 2009, New Astronomy, 14, 630
\bibitem[\protect\citeauthoryear{Gieles \& Baumgardt}{2008}]{Gieles2008}
Gieles M., Baumgardt H., 2008, MNRAS, 389, L28
\bibitem[\protect\citeauthoryear{Giersz \& Heggie}{1994}]{Giersz1994}
Giersz M., Heggie D. C., 1994, MNRAS, 268, 257
\bibitem[\protect\citeauthoryear{Giersz \& Heggie}{1996}]{Giersz1996}
Giersz M., Heggie D. C., 1996, MNRAS, 279, 1037
\bibitem[\protect\citeauthoryear{G\"urkan, Freitag \& Rasio}{2004}]{Guerkan2004}
G\"urkan A., Freitag M., Rasio F.  A., 2004, Ap. J., 604, 632
\bibitem[\protect\citeauthoryear{Habibi et al.}{2013}]{Habibi2013}
Habibi, M.; Stolte, A.; Brandner, W.; Hu\"smann, B.; Motohara, K., 2013, A\&A, 556, A2
%\bibitem[\protect\citeauthoryear{Hansen \& Phinney}{1997}]{Hansen1997}
%Hansen B. M. S., Phinney E. S., 1997, MNRAS, 291, 569
\bibitem[\protect\citeauthoryear{Harfst et al.}{2007}]{Harfst2007}
Harfst S., Gualandris A., Merritt D., et al., 2007, New Astron. 12, 357
\bibitem[\protect\citeauthoryear{Hong, Schlegel \& Grindlay}{2004}]{Hong2004} 
Hong, J., Schlegel, E. M., Grindlay, J.E., 2004, Ap. J., 614, 508 
\bibitem[\protect\citeauthoryear{Hurley, Pols \& Tout}{2000}]{Hurley2000}
Hurley J. R., Pols O. R., Tout C. A., 2000, MNRAS, 315, 543
\bibitem[\protect\citeauthoryear{Just et al.}{2009}]{Just2009}
Just A., Berczik P., Petrov M. I., Ernst A., 2009, MNRAS, 392, 969
\bibitem[\protect\citeauthoryear{Just \& Jahrei{\ss}}{2010}]{Just2010}
Just A., Jahrei\"s, H., 2010, MNRAS, 402, 461
\bibitem[\protect\citeauthoryear{Kharchenko et al.}{2009}]{Kharchenko2009}
Kharchenko N. V., Berczik P., Petrov M. I., Piskunov A. E., R\"oser S., Schilbach E.,
Scholz R.-D., 2009, A\&A, 495, 807
\bibitem[\protect\citeauthoryear{Kharchenko et al.}{2013}]{Kharchenko2013}
Kharchenko N. V., Piskunov A. E., R\"oser S., Scholz R.-D., 2013, A\&A 558, A53
\bibitem[\protect\citeauthoryear{King}{1966}]{King1966}
King I. R., 1966, AJ, 71, 64
\bibitem[\protect\citeauthoryear{Kroupa}{2001}]{Kroupa2001}
Kroupa P., 2001, MNRAS, 322, 231
\bibitem[\protect\citeauthoryear{K\"upper et al.}{2008}]{Kuepper2008} K\"upper A. H. W., Macleod A., Heggie D. C., 2008, MNRAS, 387, 1248
\bibitem[\protect\citeauthoryear{Kustaanheimo \& Stiefel}{1965}]{Kustaanheimo1965}
Kustaanheimo P. Stiefel E. L. 1965, 
J. f{\"u}r reine angewandte Mathematik, 218, 204
\bibitem[\protect\citeauthoryear{Lamers, Gieles \& Portegies Zwart}{2005a}]{Lamers2005}
Lamers H. J. G. L. M., Gieles M., Portegies Zwart S. F., 2005a, A\&A, 429, 173
\bibitem[\protect\citeauthoryear{Lamers et al.}{2005b}]{Lamers2005b}
Lamers H. J. G. L. M., Gieles M., Bastian N., Baumgardt H., Kharchenko N. V., 
Portegies Zwart S. F., 2005b, A\&A, 441, 117
\bibitem[\protect\citeauthoryear{Lamers, Baumgardt \& Gieles}{2010}]{Lamers2010}
Lamers H. J. G. L. M., Baumgardt H., Gieles M., 2010, MNRAS, 409, 305
\bibitem[\protect\citeauthoryear{Makino \& Aarseth}{1992}]{Makino1992}
Makino J., Aarseth S. J., 1992, PASJ, 44, 141
\bibitem[\protect\citeauthoryear{Markwardt}{2009}]{Markwardt2009}
Markwardt C. B., 2009, in Proc. Astronomical Data Analysis Software and Systems XVIII, Quebec, Canada, ASP Conference Series, Vol. 411, eds. D. Bohlender, P. Dowler \& D. Durand, Astronomical Society of the Pacific, San Francisco, p. 251-254 
\bibitem[\protect\citeauthoryear{Maschberger \& Kroupa}{2007}]{Maschberger2007}
Maschberger, T., Kroupa, P., 2007, MNRAS 379, 34
\bibitem[\protect\citeauthoryear{MacKay}{1990}]{MacKay1990}
MacKay R. S., 1990, Phys. Lett. A, 145, 425
\bibitem[\protect\citeauthoryear{McLachlan}{1995}]{McLachlan1995}
McLachlan R., 1995, SIAM J. Sci. Comp., 16, 151
\bibitem[\protect\citeauthoryear{Mikkola \& Tanikawa}{1999}]{Mikkola1999} 
Mikkola S., Tanikawa K., 1999a, MNRAS, 310, 745 50
\bibitem[\protect\citeauthoryear{Mikkola \& Aarseth}{2002}]{Mikkola2002} Mikkola S., Aarseth S. J., 
2002, Cel. Mech. Dyn. Astron., 84, 343
\bibitem[\protect\citeauthoryear{Miller \& Scalo}{1978}]{Miller1978}
Miller G. E., Scalo J. M., 1978, PASP, 90, 506
\bibitem[\protect\citeauthoryear{Miocchi et al.}{2013}]{Miocchi2013}
Miocchi P. et al., 2013, Ap. J., 774, 151 
\bibitem[\protect\citeauthoryear{Miyamoto \& Nagai}{1975}]{Miyamoto1975}
Miyamoto M., Nagai R., 1975, PASJ, 27, 533
\bibitem[\protect\citeauthoryear{Mor\'e}{1978}]{More1978}
Mor\'e J., 1978, in Numerical Analysis, vol. 630, ed. G. A. Watson, 
Springer Verlag, Berlin, p. 105
\bibitem[\protect\citeauthoryear{Nitadori \& Aarseth}{2012}]{Nitadori2012}
Nitadori K., Aarseth S. J., 2012, MNRAS, 424, 545
\bibitem[\protect\citeauthoryear{Pang et al.}{2013}]{Pang2013}
Pang, X., Grebel, E. K., Allison, R. J., Goodwin, S. P., Altmann, M., Harbeck, D., Moffat, A. F. J., Drissen, L., 2013, ApJ, 764, 73
\bibitem[\protect\citeauthoryear{Parmentier \& Baumgardt}{2012}]{Parmentier2012}
Pamentier, G., Baumgardt, H., 2012, MNRAS, 427, 1940
\bibitem[\protect\citeauthoryear{R\"oser et al.}{2010}]{Roeser2010}
R\"oser, S., Kharchenko, N. V., Piskunov, A. E., Schilbach, E., Scholz, R.-D., Zinnecker, H., 
2010, AN, 331, 519
\bibitem[\protect\citeauthoryear{Ross, Mennim \& Heggie}{1997}]{Ross1997}
Ross D. J., Mennim A., Heggie D. C., 1997, MNRAS, 284, 811
\bibitem[\protect\citeauthoryear{Preto \& Tremaine}{1999}]{Preto1999} Preto M., Tremaine S. 1999, AJ, 118, 2532
R\"oser, S., Kharchenko, N. V., Piskunov, A. E., Schilbach, E., 
Scholz, R.-D., Zinnecker, H., 2010, AN, 331, 519
\bibitem[\protect\citeauthoryear{Tanikawa \& Fukushige}{2005}]{Tanikawa2005}
Tanikawa A., Fukushige T., 2005, PASP, 57, 155
\bibitem[\protect\citeauthoryear{Whitehead et al.}{2013}]{Whitehead2013}
Whitehead A. J., 2013, Ap. J. 778, 118
\bibitem[\protect\citeauthoryear{Wielen}{1971}]{Wielen1971}
Wielen R., 1971, A\&A, 13, 309
\bibitem[\protect\citeauthoryear{Wielen}{1985}]{Wielen1985}
Wielen R., 1985, in: {\it Dynamics of star clusters}, Proceedings of the Symposium,
Princeton, NJ, May 29 - June 1, 1984,  Dordrecht, D. Reidel Publishing Co., 1985, p. 449
\bibitem[\protect\citeauthoryear{Yoshida}{1990}]{Yoshida1990}
Yoshida H. 1990, Phys. Lett. A, 150, 262
\end{thebibliography}
\end{document}